\documentclass[12pt, oneside, reqno]{amsart}
\usepackage{amsmath, amsthm, amssymb, amsfonts }
\usepackage{graphicx}
\usepackage{tikz}
\usepackage{tikz-cd}
\usepackage{tensor}
\usepackage[margin = 1in]{geometry}
\usepackage[numbers,sort&compress]{natbib}
\usepackage{enumitem}
\usepackage[export]{adjustbox}
\usepackage{wrapfig}
\usepackage{url}
\usepackage{hyperref}
\usepackage{braids}
\usepackage{xcolor}
\usepackage{listings}

\theoremstyle{definition}
\newtheorem{theorem}{Theorem}[section]

\newtheorem{lemma}[theorem]{Lemma}

\newtheorem{definition}[theorem]{Definition}

\newcommand{\bfSU}{\textbf{SU}}
\newcommand{\ket}[1]{|#1\rangle}
\newcommand*\circled[1]{\tikz[baseline=(char.base)]{
            \node[shape=circle,draw,inner sep=1pt] (char) {#1};}}

\definecolor{codegreen}{rgb}{0,0.6,0}
\definecolor{codegray}{rgb}{0.5,0.5,0.5}
\definecolor{codepurple}{rgb}{0.58,0,0.82}
\definecolor{backcolour}{rgb}{0.95,0.95,0.92}

\lstdefinestyle{mystyle}{
    backgroundcolor=\color{backcolour},   
    commentstyle=\color{codegreen},
    keywordstyle=\color{magenta},
    numberstyle=\tiny\color{codegray},
    stringstyle=\color{codepurple},
    basicstyle=\ttfamily\footnotesize,
    breakatwhitespace=false,         
    breaklines=true,                 
    captionpos=b,                    
    keepspaces=true,                 
    numbers=left,                    
    numbersep=5pt,                  
    showspaces=false,                
    showstringspaces=false,
    showtabs=false,                  
    tabsize=2
}
\title[]{Universal topological quantum computing via double-braiding in SU(2) Witten-Chern-Simons theory}

\author{Adrian L. Kaufmann}
\address{West Lafayette Junior/Senior High School, West Lafayette, IN 47906}
\email{kaufmannadi25@gmail.com}

\author{Shawn X. Cui}
\address{Department of Mathematics, Department of Physics and Astronomy\\Purdue University\\West Lafayette, IN\\47907\\USA}
\email{cui177@purdue.edu}

\date{December 2023}

\begin{document}
\begin{abstract}
    We study the problem of universality in the anyon model described by the $SU(2)$ Witten-Chern-Simons theory at level $k$. A classic theorem of Freedman-Larsen-Wang states that for $k \geq 3, \ k \neq 4$, braiding of the anyons of topological charge $1/2$ is universal for topological quantum computing. For the case of one qubit, we prove a stronger result that double-braiding of such anyons alone is already universal.
\end{abstract}

\maketitle
\section{Introduction}
Topological quantum computing  is an approach to building a fault-tolerant quantum computer using certain quasi-particles, called anyons, in two dimensions. The physical systems hosting anyons are 2-dimensional topological phases of matter. Topological phases are gapped quantum phases which go beyond Landau's theory of symmetry breaking and local order parameters; instead, they obey a new order called topological quantum order \cite{wen1990topological}. Such phases exhibit  several remarkable properties including robust ground state degeneracy depending on the topology of the underlying system, long-range entanglement, protected gapless edge modes, fractionalized quasi-particle excitations, and exotic exchange statistics. The robustness of the ground/excited state space provides an ideal place to store quantum information as logical qubits. Braiding of anyons, a process of adiabatically interchanging anyons, induces a unitary transformation on the state space. These unitary transformations remain unchanged under local deformations of the braiding world-lines, and hence serve as logical quantum gates. This method of encoding and manipulating information in global degrees of anyons is called topological quantum computing \cite{freedman2002modular, kitaev2003fault}, and it has the advantage of achieving fault tolerance at the `hardware' level. This is an active area of research both theoretically and experimentally. See \cite{nayak2008non, wang2010topological} for a comprehensive review. For some more recent progress on this area, see, for instance, \citep{aghaee2024interferometric, aghaee2023inas, minev2024realizing, nakamura2020direct,  xu2024non, wu2023simulating,  zhan2022universal}, and references therein.

An important family of topological phases is described by the Witten-Chern-Simons theory associated with the Lie group $SU(2)$ and a level $k \in \mathbb{Z}_{\geq 0}$ \cite{witten1989quantum}. Denote the theory by $\bfSU(2)_k$. It is among the earliest-studied anyon models. By a classic result of \cite{freedman2002two}, except for a few values of $k$, the model is universal for topological quantum computing, i.e., it is equivalent to the standard circuit model. Furthermore, the theory has various potential realizations in fractional quantum Hall (FQH) systems. For example, $\bfSU(2)_2$ contains the Ising anyon expected to exist in FQH with filling factor $\nu = 5/2$, and $\bfSU(2)_3$ contains the Fibonacci anyon expected to exist in $\nu = 12/5$ FQH.

We elaborate a bit more on the universality of $\bfSU(2)_k$. It contains an anyon type, which we denote by $\tau$, with topological charge $1/2$. Fusing two type-$\tau$ anyons produces either the ground state or a type-$\tau$ anyon.  By iteratively fusing $\tau$ with itself, every anyon type in this model can be produced. In this sense, $\tau$ is the most critical anyon type in $\bfSU(2)_k$. Denote by $V^{\tau^{\otimes 3}}_{\tau}$ the space of three $\tau$ anyons with total type equal to $\tau$, or equivalently, the space of four type-$\tau$ anyons (with total type trivial). This space has dimension two, and hence is a qubit. Braiding of the $\tau$ anyons induces a unitary representation of the braid group $B_3$ on the qubit. The theorem of \cite{freedman2002two} states that, for all $k \geq 3$, $k \neq 4, \ 8$, this representation has a dense image in $U(V^{\tau^{\otimes 3}}_{\tau})$, implying universal quantum computing on one qubit by braiding. In fact, \cite{freedman2002two} shows that for the above values of $k$ and for any $n \geq 3$, the image of the braid group representation is dense in $V^{\tau^{\otimes n}}_{\tau}$ \footnote{This result also holds for $k=8$, but $n$ needs to be at least $5$.}. These results set the theoretical foundation for universal topological quantum computing.

\begin{figure}
\begin{tikzpicture}[->,>=stealth, thick]

\begin{scope}[xshift = 0cm]
    \node[draw,circle, fill = white, white, minimum size = 1cm](node1) at (0,0) {};
    \node[draw,circle, fill = white, white,  minimum size = 1cm](node2) at (2,0) {};
    
    \fill[black] (0,0) circle[radius = 0.1cm];
    \fill[black] (2,0) circle[radius = 0.1cm];

    \draw[->] (node1)[out = -30, in = -150] to (node2);
    \draw[->] (node2)[out = 150, in = 30] to (node1);
\end{scope}

\begin{scope}[xshift = 6cm]
    \node[draw,circle, fill = white, white, minimum size = 1cm](node1) at (0,0) {};
    \node[draw,circle, fill = white, white,  minimum size = 1cm](node2) at (2,0) {};
    
    \fill[black] (0,0) circle[radius = 0.1cm];
    \fill[black] (2,0) circle[radius = 0.1cm];

    \draw[->] (node1)[out = -30, in = -90] to (2.5,0) [out = 90, in = 30] to (node1);
\end{scope}

\end{tikzpicture}
\caption{(Left) A counterclockwise braiding of two anyons; (Right) A counterclockwise double-braiding of two anyons.}\label{fig:braiding_vs_double_braiding}
\end{figure}
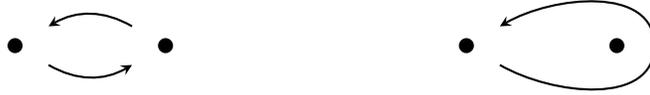

In this paper, we focus on the qubit $V^{\tau\tau\tau}_{\tau}$ and study the representation from braiding,
\begin{equation}
    \rho_k \colon B_3 \to U(V^{\tau\tau\tau}_{\tau}).
\end{equation}
Recall that the braid group on $n$ strands $B_n$ has the standard generators $\sigma_1, \cdots,\sigma_{n-1}$ (see Sec. \ref{sec:TQC_review}). $B_n$ acts on the space of $n$ identical anyons of certain type where $\sigma_i$ corresponds to a counterclockwise braiding the $i$-th and $(i+1)$-th anyon.
We call $\sigma_i^2$ a \emph{double-braiding}, and it corresponds to moving the $i$-th anyon counterclockwise around the $(i+1)$-th anyon and returning to its original position in the end. See Figure \ref{fig:braiding_vs_double_braiding}. More generally, any braid in the group generated by the $\sigma_i^2\ '$s is also called a double-braiding. Specializing to the case of one qubit $V^{\tau\tau\tau}_{\tau}$ in $\bfSU(2)_k$ with $k \geq 3$, $k \neq 4, \ 8$, while \cite{freedman2002two} states that the image of $B_3$ under $\rho_k$ is dense in $U(V^{\tau\tau\tau}_{\tau})$, we prove that the image of the subgroup of double-braidings alone is dense in $U(V^{\tau\tau\tau}_{\tau})$.

Our result is of significance in several aspects. Firstly, it is mathematically stronger than the theorem of \cite{freedman2002two} adapted to one qubit, and may reveal some hidden structure in the representations of the braid group. Secondly, although our result is stated only for one qubit, it is not too much restricted, as any entangling 2-qubit gate plus 1-qubit gates are universal for quantum computing. Thirdly, our result generalizes the work of \cite{cui2019search} where the Fibonacci anyon was shown to be universal by double-braidings, which corresponds to the case $k=3$ here. Moreover, using the double-braiding-universality of the Fibonacci anyon, the authors in \cite{cui2019search} provided an elegant, exponentially fast algorithm to produce entangling leakage-free 2-qubit gates which are necessary for universal quantum computing on multi-qubits. With the result in the current paper, the algorithm of \cite{cui2019search} can be straightforwardly adapted to other $\bfSU(2)_k$. Fourthly, from a physical point of view, double-braidings have the advantage that at each step only one anyon needs to move, it needs to move at most to the vicinity of its nearest neighbor, and it returns to its own position immediately after the move. Hence, there is no need to track the positions of the involved anyons. This approach mitigates the errors associated with the control of anyons, and simplifies the operations on them,  thereby potentially reducing the experimental challenges in realizing topological quantum computing. 
It should be noted that the group of double-braidings is strictly smaller than the pure braid group consisting of braids where anyons return to their positions \emph{eventually}. 

It is now a well-established dictionary that 2-dimensional topological phases of matter are characterized by the structure of a unitary modular tensor category which is a braided category satisfying additional conditions. A double-braiding is also called a \emph{twine} structure in the braided category introduced by \cite{bruguieres2006double}. The twine structure can be formalized and defined on non-braided categories. There exist monoidal categories without a braiding structure, but with a twine structure. An example of this is the fermionic Moore-Read fusion category \cite{liptrap2010hypergroups}. While most FQH states are expected to fit in the framework of modular tensor categories, some do not seem so. Instead, they might be described by twine fusion categories \cite{liptrap2010hypergroups, bonderson2007non}. Our work on double-braidings could provide insight into exploring the power of quantum computing in those systems.

We conjecture that our result on the universality of double-braidings also hold for the case of multi-qubits, i.e., the space of more than three anyons. For that generalization, the techniques utilized in this paper may not apply. Instead, it is possible to make use of the Lie-theoretical tools on the two-eigenvalue problem in \cite{freedman2002two}. We leave this as a future direction.

After the first version of the manuscript was posted, we were made aware\footnote{S.X.C would like to thank Sachin  Valera for pointing out the result in \cite{simon2006topological}} that Theorem \ref{thm: main} follows from a result in \cite{simon2006topological}, which proved that if braiding of $n$ identical anyons is universal, then universality is also achieved by moving only one anyon around the other $n-1$ anyons. Specializing to $n=3$, the group of braids in which only the second anyon moves is precisely the group of double braids. Then, combining the braiding universality of $\bfSU(2)_k$ of \cite{freedman2002two} and result of \cite{simon2006topological} for $n=3$, we arrive at Theorem \ref{thm: main}. However, it should be noted that both \cite{freedman2002two} and \cite{simon2006topological} relied heavily on abstract Lie-theoretical results such as the characterization of normal subgroups of Lie groups. In contrast, our proof for Theorem \ref{thm: main} is explicit and only involves elementary tools. We hope the proof itself is interesting to readers who might want to extend it to other anyon theories.

The rest of the paper is organized as follows. Section \ref{sec:TQC_review} gives a brief overview on the algebraic theory of anyons and the setup in topological quantum computing, with a more detailed explanation in Appendix \ref{sec:anyon_model_detail}. In Section \ref{subsec:SU2_data} - \ref{subsec:1_qubit_model_SU2}, we provide data for $\bfSU(2)_k$ and explicit calculations of the braid group representations. Section \ref{subsec:main_result} contains the main result.

\section{Topological quantum computing with anyons}
\label{sec:TQC_review}
In this section, we provide a very brief introduction to topological quantum computing (TQC). There are many references with more comprehensive discussions on this subject, such as \cite{wang2010topological, nayak2008non, cui2015universal, rowell2018mathematics}.

Mathematically, an anyon model is characterized by the structure of a unitary modular tensor category (MTC). An MTC can be described either in terms of abstract categorical language or by a set of  concrete data. We take the second approach and provide a partial set of data for the purpose of fixing the convention. See Appendix \ref{sec:anyon_model_detail} for a more detailed discussion of MTCs.

An anyon model has a finite set 
\begin{equation}
L = \{a,b,c, \cdots\}
\end{equation}
consisting of all the possible anyon types in a topological phase. Each anyon type $a$ has a dual anyon type $\bar{a}$. The ground state is considered as a special trivial anyon type, usually denoted by $\mathbf{1}$.  The fusion rule describes the possible outcomes when fusing two anyons. Given $a, b \in L$, we formally write,
\begin{equation}
    a \otimes b = \bigoplus_{c \in L}\ N_{ab}^c\ c,
\end{equation}
where $N_{ab}^c$ denotes the number of different channels of fusing $a$ and $b$ to result in the output $c$. If there is no way to obtain  $c$ from the fusion, then $N_{ab}^c = 0$. 
If $N_{ab}^c > 0$, we say $c$ is a \emph{total type} or \emph{total charge} of $a$ and $b$, and call the triple $(a,b;c)$ \emph{admissible}. For simplicity, in the following discussions \textbf{we will assume $N_{ab}^c$ is either $0$ or $1$}, i.e., the anyon model is multiplicity-free. 

For anyon types $c, a_1, \cdots, a_n$, denote by $V_{c}^{a_1a_2\cdots a_n}$ the space of states representing $n$ anyons $a_1, \cdots, a_n$ with total charge $c$. A basis for such a space can be described as follows. Choose an upward-growing binary tree with one root at the bottom and $n$ leaves at the top. See Figure \ref{fig:basis_n_anyons_2} for an illustration.  Label the root by $c$ and the leaves, from left to right, by $a_1, a_2, \cdots, a_n$. Now label each internal edge $e$ by an anyon type $b_e$ such that at each fork, the relevant triple of labels are admissible. Then the binary tree with all possible labels $\{b_e\}$ of internal edges form a basis of  $V_{c}^{a_1a_2\cdots a_n}$, called a \emph{splitting-tree basis}. For each labeled binary tree, one can  interpret the state it represents as a splitting process,  where and throughout the context, the time direction is from bottom to top. For example, the state represented by the tree in Figure \ref{fig:basis_n_anyons_2} is obtained by splitting $c$ into $b_{n-2}$ and $a_n$, followed by splitting $b_{n-2}$ into $b_{n-3}$ and $a_{n-1}$, $\cdots\cdots$, followed by splitting $b_1$ into $b_0 = a_1$ and $a_2$. 

\begin{figure}
\begin{tikzpicture}
\draw[thick] (0.5,-0.5) -- (-1,1)node[above]{$a_1$};
\draw[thick] (0,1)node[above]{$a_2$} -- (-0.5,0.5)node[below]{$b_1$};
\draw[thick] (0,0)node[below]{$b_2$} -- (1,1)node[above]{$a_3$};
\draw[thick] (2,1)node[above]{$a_4$} -- (0.5,-0.5);
\draw (3,1) node{$\cdots$};
\draw[thick] (4,1)node[above]{$a_n$} -- (1.5,-1.5);

\draw[thick] (0.5,-0.5) -- (2,-2)node[below]{$c$};
\fill[white] (1,-1) circle[radius = 0.3cm];
\draw (1,-1) node[rotate = -45]{$\cdots$};
\end{tikzpicture}
\caption{A basis of $V_{c}^{a_1a_2\cdots a_n}$ corresponding to a binary tree}\label{fig:basis_n_anyons_2}
\end{figure}

For $n=3$, there are two splitting trees, with one internal edge, as shown on both sides of the equation below. The basis corresponding to the tree on the left side of the equation consists of all possible labelings $m$ of the internal edge so that $(a,b;m)$ and $(m,c;d)$ are both admissible. Similarly, the basis for the tree on the right side consists of labelings $n$ of the internal edge so that $(b,c;n)$ and $(a,n;d)$ are both admissible. Denote the matrix change between the two bases by $F_{d}^{abc}$. More explicitly,
\begin{equation}
\begin{aligned}
\begin{tikzpicture}
\begin{scope}
\draw[thick] (0.5,-0.5)node[below]{$d$} --(0,0) -- (-1,1)node[above]{$a$};
\draw[thick] (0,0) -- (1,1)node[above]{$c$};
\draw[thick] (0,1)node[above]{$b$} -- (-0.5,0.5)node[below]{$m$};
\end{scope}

\draw (2,0.25) node{$= \, \sum\limits_{n} \, F^{abc}_{d;nm}$};

\begin{scope}[xshift = 4cm]
 \draw[thick] (0,0) -- (-1,1)node[above]{$a$};
 \draw[thick] (-0.5,-0.5)node[below]{$d$} --(0,0) -- (1,1)node[above]{$c$};
 \draw[thick] (0,1)node[above]{$b$} -- (0.5,0.5)node[below]{$n$};
\end{scope}
\end{tikzpicture}
\end{aligned}
\end{equation}
where $F^{abc}_{d;nm}$ is the $(n,m)$-entry of $F_{d}^{abc}$, and the sum is over all labelings $n$ as described above. Note that, here the anyon types $n$ and $m$ are used as the indices of the entries of $F_{d}^{abc}$.  We call $F_{d}^{abc}$ an $F$-matrix, its entries $F$-symbols or $6j$-symbols, and the identity in the above equation an $F$-move. 

The process of swapping positions of anyons is called a braiding.  A braiding induces a unitary transformation on the state space. Consider two anyons $a$ and $b$ with total type $c$. A counterclockwise braiding of $a$ and $b$ maps a state in $V_{c}^{ab}$ to one  in $V_{c}^{ba}$. Since both spaces have dimension one, there exists a phase $R^{ba}_c$ such that the following equality holds,
\begin{equation}
\begin{aligned}
\begin{tikzpicture}[thick]
\begin{scope}
\draw (0,0) -- (-0.5,0.5)node[left]{$a$};
\draw (0,0) -- (0.5,0.5)node[right]{$b$};
\draw (0,0) -- (0,-0.5)node[below]{$c$};
\braid  (br) at (-0.5, 2) s_1^{-1};
\draw (br-1-s) node[left]{$b$};
\draw (br-2-s) node[right]{$a$};
\end{scope}

\draw (1.5,0.5) node{$=$};
\draw (2.5,0.5) node{$R^{ba}_{c}$};

\begin{scope}[xshift = 3.5cm]
\draw (0,0) -- (-0.5,0.5) -- (-0.5, 2)node[left]{$b$};
\draw (0,0) -- (0.5,0.5) -- (0.5, 2)node[right]{$a$};
\draw (0,0) -- (0,-0.5)node[below]{$c$};
\end{scope}
\end{tikzpicture}
\end{aligned}
\end{equation}
The above equality is called an $R$-move, and $R^{ba}_c$ is called an $R$-symbol. The set of $F$- and $R$-symbols are crucial for calculations in the anyon model.

Now, we discuss how to perform quantum computing with anyons. The state space of multi-anyons is the logical space to store information. Typically, one chooses multiple identical anyons, say $n$ type-$a$ anyons with total type $c$ for some $n$. Denote this space by $V_{c}^{a^{\otimes n}}:= V_{c}^{aa\cdots a}$. The type $a$ needs to be non-Abelian so that the dimension of $V_{c}^{a^{\otimes n}}$ approaches to infinity as $n \to \infty$. In general, $V_{c}^{a^{\otimes n}}$ does not have a natural tensor product structure, and we need to choose a subspace which does have a tensor product structure as multi-qubits or multi-qudits. The computational basis for the qudits can be chosen to be any splitting-tree basis.

\begin{figure}
\begin{tikzpicture}
\braid[number of strands = 4, thick](braid1) at (0,0) s_2^{-1};
\draw (0.5, -0.75) node{$\cdots$};
\draw (2.5, -0.75) node{$\cdots$};
\node[at = (braid1-1-e), below]{$1$};
\node[at = (braid1-2-e), below]{$i+1$};
\node[at = (braid1-3-e), below]{$i$};
\node[at = (braid1-4-e), below]{$n$};
\begin{scope}[xshift = 7cm]
\braid[number of strands = 4, thick](braid1) at (0,0) s_2;
\draw (0.5, -0.75) node{$\cdots$};
\draw (2.5, -0.75) node{$\cdots$};
\node[at = (braid1-1-e), below]{$1$};
\node[at = (braid1-2-e), below]{$i+1$};
\node[at = (braid1-3-e), below]{$i$};
\node[at = (braid1-4-e), below]{$n$};
\end{scope}
\end{tikzpicture}
\caption{(Left) the braid diagram $\sigma_i$; (Right) the braid diagram $\sigma_i^{-1}$.}\label{fig:braid_generator}
\end{figure}
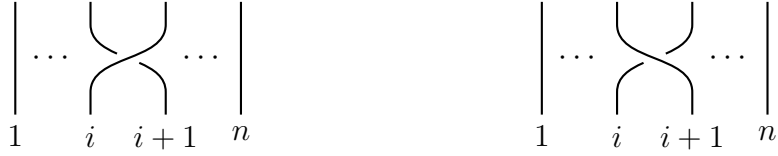

Braiding of anyons induces a representation of the braid group and acts as quantum gates on the logical space $V_{c}^{a^{\otimes n}}$. Recall that the $n$-strand braid group $B_n$ has a presentation as
\begin{align}
B_n = \langle \sigma_1, \cdots, \sigma_{n-1}\,|\, &\sigma_{i}\sigma_{i+1}\sigma_{i} = \sigma_{i+1}\sigma_{i}\sigma_{i+1}, \nonumber \\
&\sigma_{i}\sigma_{j} = \sigma_{j}\sigma_{i}, |i-j|>1\, \rangle,
\end{align}
where $\sigma_i$ (resp. $\sigma_i^{-1}$) corresponds to the braid diagram in Figure \ref{fig:braid_generator}(Left) (resp. (Right) ). 

We have a representation,
\begin{equation}
    \rho\colon B_n \to U(V_{c}^{a^{\otimes n}}),
\end{equation}
where $\sigma_i$ (resp. $\sigma_i^{-1}$) acts on $V_{c}^{a^{\otimes n}}$ by counterclockwise (resp. clockwise) braiding of the $i$-th with the $(i+1)$-th anyon. With a chosen splitting-tree basis, a matrix for each braid generator can be computed using $F$- and $R$-symbols. The set of quantum gates obtained from braiding is the image of this braid group representation.

\section{Universality in $\bfSU(2)_k$ anyons}\label{sec:universality_SU2}

\subsection{$\bfSU(2)_k$ anyons}\label{subsec:SU2_data}
Recall that the $SU(2)$  Witten-Chern-Simons theory at level $k \in \mathbb{Z}_{\geq 0}$ is denoted by $\bfSU(2)_k$. The MTC corresponding to $\bfSU(2)_k$ is constructed from finite-dimensional representations of the quantum group $U_q(\text{sl}_2)$ for $q = e^{\frac{2\pi i}{k+2}}$. It also describes the Wess–Zumino–Witten conformal field theory \cite{wess1971consequences, witten1983global}.

Below we describe the data of the $\bfSU(2)_k$ MTC obtained from Sec. 5.4 of \cite{bonderson2007non}, though the data also appeared in earlier literature in different forms. The MTC has $k+1$ anyon types   (i.e. simple objects) labeled by half-integers $0, \frac{1}{2}, \frac{2}{2}, \dots, \frac{k}{2}$, where $0$ denotes the trivial anyon type. The fusion rule is given by,
\begin{equation}
    j_1\otimes j_2=\bigoplus_{j=|j_1-j_2|}^{\min\{j_1+j_2,k-j_1-j_2\}} j,
\end{equation}
where $j$ has an increment of 1 in the above sum, implying that for admissible $(j_1, j_2;j)$, $j_1 + j_2 + j$ is always an integer. Throughout the context, fix $q = e^{\frac{2\pi i}{k+2}}$, and denote by $q^x:= e^{\frac{2\pi i x}{k+2}}$. For $n \in \mathbb{Z}_{\geq 0} $, the quantum integer $[n]$ and the quantum factorial are defined by,
\begin{align}
    [n]_q:= \frac{q^{\frac{n}{2}} - q^{-\frac{n}{2}}}{q^{\frac{1}{2}} - q^{-\frac{1}{2}}}, \quad\quad [n]_q! = ([n-1]_q!)\ [n], \quad\quad [0]_q! = 1.
\end{align}
The $R$-symbols are given by,
\begin{equation}
    R_j^{j_1,j_2}=(-1)^{j-j_1-j_2}q^{\frac{1}{2}(j(j+1)-j_1(j_1+1)-j_2(j_2+1))}.
\end{equation}
The $F$-symbols are given by,
\begin{equation}
    F_{j;j_{12},j_{23}}^{j_1,j_2,j_3} = [F_j^{j_1,j_2,j_3}]_{j_{12},j_{23}}=(-1)^{j_1+j_2+j_3+j} \sqrt{[2j_{12}+1]_q [2j_{23}+1]_q} \left\{\begin{matrix}
        j_1&j_2&j_{12}\\j_3&j&j_{23}
    \end{matrix}\right\}_q,
\end{equation}
where
\begin{eqnarray}
    \left\{\begin{matrix}
        j_1&j_2&j_{12}\\j_3&j&j_{23}
    \end{matrix}\right\}_q=\Delta(j_1,j_2,j_{12}) \Delta(j_{12},j_3,j) \Delta(j_2,j_3,j_{23}) \Delta(j_1,j_{23},j)\times\nonumber\\
    \sum_z\Bigg\{\frac{(-1)^z[z+1]_q!}{[z-j_1-j_2-j_{12}]_q![z-j_{12}-j_3-j]_q![z-j_2-j_3-j_{23}]_q![z-j_1-j_{23}-j]_q!}\\\times\frac{1}{[j_1+j_2+j_3+j-z]_q![j_1+j_{12}+j_3+j_{23}-z]_q![j_2+j_{12}+j+j_{23}-z]}\Bigg\},\nonumber
\end{eqnarray}
where the sum is over $z$ with an increment of $1$ for which all the $[\cdot]_q!$ in the sum are defined,   and 
\begin{equation}
    \Delta(j_1,j_2,j_3)=\sqrt{\frac{[-j_1+j_2+j_3]_q![j_1-j_2+j_3]_q![j_1+j_2-j_3]}{[j_1+j_2+j_3+1]_q!}}.
\end{equation}

A fact that will not be used in this paper is that a close cousin of the $\bfSU(2)_k$ MTC is the Temperley-Lieb-Jones MTC obtained from skein theory \cite{turaev2010quantum}. Under a proper translation between the level $k$ and the Kauffman variable $A$ in the skein theory, the $\bfSU(2)_k$ MTC and the Temperley-Lieb-Jones MTC are equivalent as braided fusion categories, but differ by a ribbon twist. 

\subsection{The 1-qubit model}\label{subsec:1_qubit_model_SU2}
For $k \geq 2$, we consider the anyon type labeled by $\tau:= \frac{1}{2}$ in the $\bfSU(2)_k$ model.  For $k = 2$, $\tau$ is the Ising anyon, while for $k = 3$, $\tau$ is closely related to the Fibonacci anyon \footnote{The anyon of type $\frac{1}{2}$ in $\bfSU(2)_3$ is the composite of the Fibonacci anyon with a semion \cite{bonderson2007non}.}. 

\begin{figure}
\begin{tikzpicture}[thick]
\begin{scope}
\draw[thick] (0.5,-0.5)node[below]{$\tau$} --(0,0) -- (-1,1)node[above]{$\tau$};
\draw[thick] (0,0) -- (1,1)node[above]{$\tau$};
\draw[thick] (0,1)node[above]{$\tau$} -- (-0.5,0.5)node[below]{$a$};
\end{scope}

\begin{scope}[xshift = 5cm]
\draw (0,0) -- (-1.5,1.5)node[above]{$\tau$};
\draw (-1, 1)node[below]{$b$} -- (-0.5,1.5)node[above]{$\tau$};
\draw (1, 1)node[below]{$b$} -- (0.5,1.5)node[above]{$\tau$};
\draw (0,0) -- (1.5,1.5)node[above]{$\tau$};
\draw (0,0)-- (0,-0.5)node[below]{$0$};
\end{scope}
\end{tikzpicture}
\caption{(Left) A splitting-tree basis for $V_{\tau}^{\tau\tau\tau}$; (Right) A splitting-tree basis for $V_{0}^{\tau\tau\tau\tau}$, where $\tau$ is the anyon of type $\frac{1}{2}$. $a,b = 0,1$.}\label{fig:1_qubit_basis}
\end{figure}
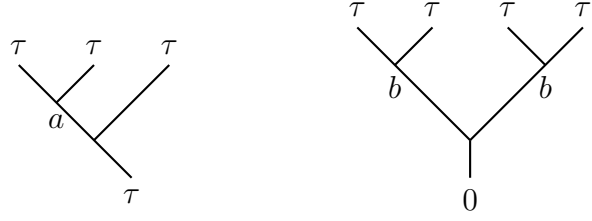

There are two standard ways to obtain a qubit using $\tau$ anyons. They are the dense encoding and the sparse encoding corresponding to the spaces $V^{\tau\tau\tau}_{\tau}$ and $V^{\tau\tau\tau\tau}_{0}$, respectively. That is, the dense encoding takes, as a qubit, the space of three $\tau$ anyons with total type $\tau$, while the sparse encoding takes the space of four $\tau$ anyons with total type $0$. From the fusion rule,
\begin{align}
    \frac{1}{2} \otimes \frac{1}{2} = 0 \oplus 1, \quad \frac{1}{2} \otimes 1           = \frac{1}{2} \oplus \frac{3}{2}, \quad 0           \otimes j           = j, 
\end{align}
both $V^{\tau\tau\tau}_{\tau}$ and $V^{\tau\tau\tau\tau}_{0}$ are of two dimensions with a splitting-tree basis given in Figure \ref{fig:1_qubit_basis}. Denote the splitting-tree basis in Figure \ref{fig:1_qubit_basis} (Left) by $\{\ket{x}: \ x = 0, 1\}$ and the one in Figure \ref{fig:1_qubit_basis} (Right) by $\{\ket{x}': \ x = 0, 1\}$.

The braiding of the $\tau$ anyons induces representations of the braid groups with the action of the generator $\sigma_i$ given by the counter-clockwise swap of the $i$-th and the $(i+1)$-th anyon,
\begin{align}
    \rho_k \colon &B_3 \to U(V^{\tau\tau\tau}_{\tau}),\\
    \rho_k'\colon &B_4 \to U(V^{\tau\tau\tau\tau}_{0}).
\end{align}

For the dense encoding, under the basis $\{\ket{x}: \ x = 0, 1\}$ of $V^{\tau\tau\tau}_{\tau}$, the action of the generators $\sigma_1$ and $\sigma_2$ are computed in Figure \ref{fig:sigma_1_action} and Figure \ref{fig:sigma_2_action}, respectively.

\begin{figure}
\begin{tikzpicture}[thick]
\begin{scope}[xshift = 0cm, yshift = 0cm]
\draw (0,-0.5)node[below]{$\tau$} --(0,0) -- (-1,1);
\draw (0,0) -- (1,1);
\draw (0,1) -- (-0.5,0.5)node[below]{$x$};
\braid [number of strands = 3] (br) at (-1, 2.5) s_1^{-1};
\draw (br-1-s)node[right]{$\tau$};
\draw (br-2-s)node[right]{$\tau$};
\draw (br-3-s)node[right]{$\tau$};
\end{scope}

\draw (3,1) node{$= \ R^{\tau\tau}_{x}$};

\begin{scope}[xshift = 6cm, yshift = 0cm]
\draw (0,-0.5)node[below]{$\tau$} --(0,0) -- (-1,1);
\draw (0,0) -- (1,1);
\draw (0,1) -- (-0.5,0.5)node[below]{$x$};
\braid [number of strands = 3, height = 1 cm] (br) at (-1, 2.5) 1;
\draw (br-1-s)node[right]{$\tau$};
\draw (br-2-s)node[right]{$\tau$};
\draw (br-3-s)node[right]{$\tau$};
\end{scope}
\end{tikzpicture}
\caption{The action of $\sigma_1$ on $V_{\tau}^{\tau\tau\tau}$.}\label{fig:sigma_1_action}
\end{figure}

\begin{figure}
\begin{tikzpicture}[thick]
\begin{scope}[xshift = 0cm, yshift = 0cm]
\draw (0,-0.5)node[below]{$\tau$} --(0,0) -- (-1,1);
\draw (0,0) -- (1,1);
\draw (0,1) -- (-0.5,0.5)node[below]{$x$};
\braid [number of strands = 3] (br) at (-1, 2.5) s_2^{-1};
\draw (br-1-s)node[right]{$\tau$};
\draw (br-2-s)node[right]{$\tau$};
\draw (br-3-s)node[right]{$\tau$};
\end{scope}

\draw (2,1) node{$=$};
\draw (5,1) node{$\sum\limits_{y \in \{0,1\}} F^{\tau\tau\tau}_{\tau;yx} $};

\begin{scope}[xshift = 9cm, yshift = 0cm]
\draw (0,-0.5)node[below]{$\tau$} --(0,0) -- (-1,1);
\draw (0,0) -- (1,1);
\draw (0,1) -- (0.5,0.5)node[below]{$y$};
\braid [number of strands = 3] (br) at (-1, 2.5) s_2^{-1};
\draw (br-1-s)node[right]{$\tau$};
\draw (br-2-s)node[right]{$\tau$};
\draw (br-3-s)node[right]{$\tau$};
\end{scope}

\begin{scope}[xshift = 0cm, yshift = -4cm]
\draw (-4.5,1) node{$=$};
\draw (-3,1) node{$\sum\limits_{y \in \{0,1\}} F^{\tau\tau\tau}_{\tau;yx} R^{\tau\tau}_y$};

\draw (0,-0.5)node[below]{$\tau$} --(0,0) -- (-1,1);
\draw (0,0) -- (1,1);
\draw (0,1) -- (0.5,0.5)node[below]{$y$};
\braid [number of strands = 3, height = 1 cm] (br) at (-1, 2.5) 1;
\draw (br-1-s)node[right]{$\tau$};
\draw (br-2-s)node[right]{$\tau$};
\draw (br-3-s)node[right]{$\tau$};

\draw (2,1) node{$=$};
\draw (5,1) node{$\sum\limits_{y,z \in \{0,1\}} F^{\tau\tau\tau}_{\tau;yx}R^{\tau\tau}_{y}(F^{\tau\tau\tau}_{\tau})^{-1}_{zy} $};
\end{scope}

\begin{scope}[xshift = 9cm, yshift = -4cm]
\draw (0,-0.5)node[below]{$\tau$} --(0,0) -- (-1,1);
\draw (0,0) -- (1,1);
\draw (0,1) -- (-0.5,0.5)node[below]{$z$};
\braid [number of strands = 3, height = 1 cm] (br) at (-1, 2.5) 1;
\draw (br-1-s)node[right]{$\tau$};
\draw (br-2-s)node[right]{$\tau$};
\draw (br-3-s)node[right]{$\tau$};
\end{scope}

\end{tikzpicture}
\caption{The action of $\sigma_2$ on $V_{\tau}^{\tau\tau\tau}$. The first and the third equality are due to an $F$-move and an inverse $F$-move, respectively. The second equality is due to an $R$-move.}\label{fig:sigma_2_action}
\end{figure}

Hence,
\begin{align}
    \rho_k(\sigma_1)\ket{x} = R^{\tau\tau}_{x} \ket{x}
\end{align}
\begin{align}
    \rho_k(\sigma_2)\ket{x} = \sum_{y,z \in \{0,1\}}\  F^{\tau\tau\tau}_{\tau;yx}\ R^{\tau\tau}_{y}\ (F^{\tau\tau\tau}_{\tau})^{-1}_{zy}\ \ket{z}.
\end{align}
Denote by,
\begin{equation}
    R = 
    \begin{pmatrix}
        R^{\tau\tau}_{0} & 0 \\
        0                & R^{\tau\tau}_{1}\\
    \end{pmatrix},
    \quad\quad
    F = F^{\tau\tau\tau}_{\tau} = 
    \begin{pmatrix}
        F^{\tau\tau\tau}_{\tau;00} &  F^{\tau\tau\tau}_{\tau;01}\\
        F^{\tau\tau\tau}_{\tau;10} &  F^{\tau\tau\tau}_{\tau;11}\\
    \end{pmatrix}.
\end{equation}
From the data in Section \ref{subsec:SU2_data}, 
\begin{equation}\label{eqn:R_F_unnormalized}
    R = 
    \begin{pmatrix}
        -q^{-\frac{3}{4}} & 0 \\
        0                & q^{\frac{1}{4}}\\
    \end{pmatrix},
    \quad\quad
    F = F^{\tau\tau\tau}_{\tau} = \frac{\sqrt{q}}{q+1}
    \left(
\begin{array}{cc}
 -1 & \sqrt{q+\frac{1}{q}+1} \\
 \sqrt{q+\frac{1}{q}+1} & 1 \\
\end{array}
\right).
\end{equation}
Note that $F$ is a symmetric, real, involutory matrix.
Then we have
\begin{align}
    \rho_k(\sigma_1) = R =
     \begin{pmatrix}
        -q^{-\frac{3}{4}} & 0 \\
        0                & q^{\frac{1}{4}}\\
    \end{pmatrix}, 
\end{align}
\begin{equation}
  \rho_k(\sigma_2) = F^{-1}RF = 
  \frac{q^{\frac{1}{4}}}{1+q}
    \left(
\begin{array}{cc}
 q & \sqrt{q+\frac{1}{q}+1} \\
 \sqrt{q+\frac{1}{q}+1} & -\frac{1}{q} \\
\end{array}
\right).
\end{equation}

For the sparse encoding, we show in fact the image $\rho_k'(B_4)$ is the same as $\rho_k(B_3)$ if we identify $\ket{x}' \in V^{\tau\tau\tau\tau}_{0}$ with $\ket{x} \in V^{\tau\tau\tau}_{\tau}$. Under this identification, it is clear that $\rho_k'(\sigma_1) = \rho_k'(\sigma_3) = \rho_k(\sigma_1)$, while $\rho_k'(\sigma_2)$ can be computed, similar to $\rho_k(\sigma_2)$ in Figure \ref{fig:sigma_2_action}, as,
\begin{align}
    \rho_k'(\sigma_2)\ket{x}' = (F^{x\tau\tau}_{0})^{-1}_{\tau x} \ \left(\sum_{y,z \in \{0,1\}}\  F^{\tau\tau\tau}_{\tau;yx}\ R^{\tau\tau}_{y}\ (F^{\tau\tau\tau}_{\tau})^{-1}_{zy}\right)\  F^{z\tau\tau}_{0;z\tau}\ \ket{z}'.
\end{align}
From the $F$-symbols of $\bfSU(2)_k$ in Section \ref{subsec:SU2_data}, it can be checked that $F^{j_1j_2j_3}_{j; j_{23}j_{12}} = 1$ whenever $j = 0$ and the involved labels are admissible. Hence we have  $\rho_k'(\sigma_2) = \rho_k(\sigma_2)$. This shows the equality of $\rho_k'(B_4)$ and $\rho_k(B_3)$.

Therefore, from now on, we will focus on the dense encoding only. A logical qubit is given by the space $V^{\tau\tau\tau}_{\tau}$ whose computational basis is $\{\ket{0}, \ket{1}\}$ as shown in Figure \ref{fig:1_qubit_basis} (Left). The set of 1-qubit logical gates obtained from anyon braidings correspond to elements in the image $\rho_k(B_3)$. Since quantum gates are well defined only up to global $U(1)$ phases, the gates in $\rho_k(B_3)$ can be multiplied by any phase, or they should be considered as elements of the projective unitary group $PU(V^{\tau\tau\tau}_{\tau})$.

\begin{definition}
    Let $V$ be a Hilbert space, and $\mathcal{G}$ be a subset of the unitary group $U(V)$. $\mathcal{G}$ is said to be universal on $V$ if $\mathcal{G} \cup U(1)$ generate a dense subgroup of $U(V)$.
\end{definition}

To study the universality of the braiding gates, it is convenient to normalize the generators of $B_3$ so that the image of $\rho_k$ lies in $SU(V^{\tau\tau\tau}_{\tau})$. Explicitly, multiplying $-i q^{1/4}$ to the generators, we obtain the normalized representation $\tilde{\rho}_k: B_3 \to SU(V^{\tau\tau\tau}_{\tau})$,
\begin{equation}\label{eqn:R_normalized}
    \tilde{\rho}_k(\sigma_1) = \tilde{R} = 
    \begin{pmatrix}
        i\ q^{-\frac{1}{2}} & 0 \\
        0                & -i\ q^{\frac{1}{2}}\\
    \end{pmatrix},
\end{equation}
\begin{equation}\label{eqn:sigma_2_normalized}
\tilde{\rho}_k(\sigma_2) = F^{-1} \tilde{R} F = \frac{i\sqrt{q}}{q+1}
    \left(
\begin{array}{cc}
 -q & -\sqrt{q+\frac{1}{q}+1} \\
 -\sqrt{q+\frac{1}{q}+1} & \frac{1}{q} \\
\end{array}
\right).
\end{equation}
Since $\rho_k$ equals  $\tilde{\rho}_k$ up to a global phase, we will use $\rho_k$  and $\tilde{\rho}_k$ interchangeably. 

As an example, when $k = 2$, $\tilde{\rho}_2(\sigma_1)$ and $\tilde{\rho}_k(\sigma_2)$ are given, respectively, by,
\begin{equation}
    e^{\frac{\pi i}{4}}\ \left(
\begin{array}{cc}
 1 & 0 \\
 0 & -i \\
\end{array}
\right)
\quad
\text{and}
\quad
\frac{1}{\sqrt{2}}\ \left(
\begin{array}{cc}
 1 & -i \\
 -i & 1 \\
\end{array}
\right),
\end{equation}
which, together with $U(1)$ phases, generate the 1-qubit Clifford group.

The multi-qubit  models can be obtained by increasing the number of anyons utilized. We will not discuss that direction. The universality of braiding the $\tau$ anyons is a classic problem settled in \cite{freedman2002two}. We rephrase Theorem 4.1 of \cite{freedman2002two} in the setup of one qubit.
\begin{theorem}[\cite{freedman2002two}]\label{thm:universal_original}
    For any integer $k \geq 3, k \neq 4, 8$, let $V^{\tau\tau\tau}_{\tau}$ be the 1-qubit space in the $\bfSU(2)_k$ model. Then set of braiding gates corresponding to the image of the representation $\tilde{\rho}_k$ defined in Equations \ref{eqn:R_normalized} \ref{eqn:sigma_2_normalized}  is universal on $V^{\tau\tau\tau}_{\tau}$.
\end{theorem}


\subsection{Universality of double-braiding in $\bfSU(2)_k$}\label{subsec:main_result}
In this section, we prove a stronger result than Theorem \ref{thm:universal_original}. That is, for those values of $k$ in Theorem \ref{thm:universal_original}, we show that the set of double-braiding gates on $V^{\tau\tau\tau}_{\tau}$ is universal in the $\bfSU(2)_k$ model. By a double-braiding is meant a braid generated by even powers of the standard generators $\sigma_i'$s of the braid group. We will rely on two critical results. The content of Lemma \ref{lem:non-commuting} below first appeared in the work of Kitaev \cite{kitaev1997quantum} where he used the lemma to prove the universality of certain gate sets. The lemma can also be found in Section 6 of  \cite{aharonov2008fault}. 

\begin{lemma}[\cite{kitaev1997quantum}]\label{lem:non-commuting}
    Let $A$ and $B$ be two non-commuting elements of $SU(2)$ both of which are of infinite order. Then $A$ and $B$ generate a dense subgroup of $SU(2)$.
    \begin{proof}
        (\emph{sketch}.) Let $G$ be the closure of the subgroup generated by $A$ and $B$. By the assumption in the lemma, $G$ is a closed connected non-commutative Lie subgroup of $SU(2)$. Hence the Lie algebra $\mathfrak{g}$ of $G$ is a non-commutative Lie subalgebra of $\mathfrak{su}(2)$. The only such Lie subalgebra of $\mathfrak{su}(2)$ is itself. Therefore $G = SU(2)$.
    \end{proof}
\end{lemma}

\begin{theorem}[Theorem 7, \cite{conway1976trigonometric}]\label{thm:cosin_sum}
    Suppose we have at most four distinct rational multiples of $\pi$ lying strictly between $0$ and $\frac{\pi}{2}$ for which some rational linear combination of their cosines is rational but no proper subset has this property. Then the appropriate linear combination is proportional to one from the following list:
   \begin{itemize}

\item $\cos{\pi/3}=\frac{1}{3}$,
    \item $-\cos{\phi}+\cos{(\pi/3-\phi)}+\cos{(\pi/3+\phi)}=0     \;              (0<\phi<\pi/6)$,
    \item $\cos{\pi/5}-\cos{2\pi/5}=\frac{1}{2}$,
   \item $\cos{\pi/7}-\cos{2\pi/7}+\cos{3\pi/7}=\frac{1}{2}$,
   \item $\cos{\pi/5}-\cos{\pi/15}+\cos{4\pi/15}=\frac{1}{2}$,
   \item $-\cos{2\pi/5}+\cos{2\pi/15}-\cos{7\pi/15}=\frac{1}{2}$,
   \item $\cos{\pi/7}+\cos{3\pi/7}-\cos{\pi/21}+\cos{8\pi/21}=\frac{1}{2}$,
   \item $\cos{\pi/7}-\cos{2\pi/7}+\cos{2\pi/21}-\cos{5\pi/21}=\frac{1}{2}$,
   \item $-\cos{2\pi/7}+\cos{3\pi/7}+\cos{4\pi/21}+\cos{10\pi/21}=\frac{1}{2}$,
   \item $-\cos{\pi/15}+\cos{2\pi/15}+\cos{4\pi/15}-\cos{7\pi/15}=\frac{1}{2}$.
   
\end{itemize}
    
\end{theorem}

The following is the main result of the paper.
\begin{theorem}\label{thm: main}
    For any integer $k \geq 3, k \neq 4, 8$, let $\tau $ be the anyon of type $\frac{1}{2}$ in the anyon model $\bfSU(2)_k$, and  $\rho_k: B_3 \to U(V^{\tau\tau\tau}_{\tau})$ be the representation of $B_3$ on the dense-encoding 1-qubit  $V^{\tau\tau\tau}_{\tau}$. Then the images of $\sigma_1^2$ and $\sigma_2^2$ under $\rho_k$, together with phases, generate a dense subgroup of $U(V^{\tau\tau\tau}_{\tau})$. That is, the double-braiding gates alone are universal.
    \begin{proof}
    It suffices to show, for the normalized $\tilde{\rho}_k: B_3 \to U(V^{\tau\tau\tau}_{\tau})$ defined in Equations \ref{eqn:R_normalized} \ref{eqn:sigma_2_normalized}, $\tilde{\rho}_k(\sigma_1^2)$ and $\tilde{\rho}_k(\sigma_2^2)$ generate a dense subgroup of $SU(V^{\tau\tau\tau}_{\tau})$.  Let
        \begin{equation}
            A := \tilde{\rho}_k(\sigma_1^2\sigma_2^4) = \tilde{R}^2 F^{-1} \tilde{R}^4 F, \quad B := \tilde{\rho}_k(\sigma_1^2\sigma_2^6) = \tilde{R}^2 F^{-1} \tilde{R}^6 F.
        \end{equation}
        From the expressions of $\tilde{R}$ (Equation \ref{eqn:R_normalized}) and $F$ (Equation \ref{eqn:R_F_unnormalized}), the matrices of $A$, $B$, and $W:= ABA^{-1}B^{-1}$ can be calculated as (See Appendix \ref{sec:Mathematica} for a Mathematica implementation),
     
        \begin{equation}
            A = 
            \left(
\begin{array}{cc}
 -\frac{q^4+q^2-q+1}{q^3+q^2} & -\frac{\sqrt{q+\frac{1}{q}+1}
   \left(q^3-q^2+q-1\right)}{q^2 (q+1)} \\
 -\frac{\sqrt{q+\frac{1}{q}+1} \left(q^3-q^2+q-1\right)}{q+1} & -\frac{q^5+q^2+q+1}{q
   (q+1)^2} \\
\end{array}
\right)
        \end{equation}
        \begin{equation}
            B = 
            \left(
\begin{array}{cc}
 \frac{q^7+q^6+q^5+1}{q^3 (q+1)^2} & \frac{\sqrt{q+\frac{1}{q}+1}
   \left(q^5-q^4+q^3-q^2+q-1\right)}{q^3 (q+1)} \\
 \frac{\sqrt{q+\frac{1}{q}+1} \left(q^5-q^4+q^3-q^2+q-1\right)}{q (q+1)} &
   \frac{q^6+q+\frac{1}{q}+1}{q (q+1)^2} \\
\end{array}
\right)
        \end{equation}
         \begin{align}
 W_{11} &=\frac{q^{13}\! - \!3 q^{12}\! + \!6 q^{11}\!\! - \!\!11 q^{10}\! + \!16 q^9\!\! -\! \!19 q^8\! + \!22 q^7\!\! - \!\!21 q^6\! + \!19 q^5\! \!-\! \!14 q^4\! + \!10
   q^3\! - \!6 q^2\! + \!3 q\! - \!1}{q^6 (q\! + \!1)}\\
W_{12} &= \! - \!\frac{(q\! - \!1)^2 \sqrt{q\! + \!\frac{1}{q}\! + \!1} \left(q^{10}\! - \!q^9\! + \!3 q^8\! - \!4 q^7\! + \!5 q^6\! - \!6 q^5\! + \!6 q^4\! - \!5
   q^3\! + \!4 q^2\! - \!2 q\! + \!1\right)}{q^7 (q\! + \!1)}\\
W_{21} &= \frac{(q\! - \!1)^2 \sqrt{q\! + \!\frac{1}{q}\! + \!1} \left(q^{10}\! - \!2 q^9\! + \!4 q^8\! - \!5 q^7\! + \!6 q^6\! - \!6 q^5\! + \!5 q^4\! - \!4
   q^3\! + \!3 q^2\! - \!q\! + \!1\right)}{q^4 (q\! + \!1)}\\
W_{22} &= \frac{\! - \!q^{13}\! + \!3 q^{12}\!\! - \!\!6 q^{11}\! + \!10 q^{10}\!\! -\! \!14 q^9\! + \!19 q^8\!\! - \!\!21 q^7\! + \!22 q^6\!\! -\! \!19 q^5\! + \!16 q^4\!\! - \!\!11
   q^3\! + \!6 q^2\! \!- \!\!3 q\! + \!1}{q^7\! + \!q^6}
        \end{align}
We will show that, for the values of $k$ in the statement of the theorem, $A$ and $B$ are both of infinite order and they do not commute. Then by Lemma \ref{lem:non-commuting}, $A$ and $B$ generate a dense subgroup of $SU(2)$, implying the validity of the theorem. The rest of the proof is devoted to verifying the assumptions on $A$ and $B$ mentioned above.

\vspace{0.2cm}
\noindent\textbf{Proving $\mathbf{A}$ has infinite order.} 

Denote the eigenvalues of $A$ by $e^{\pm \theta_k i}$, $0 \leq \theta_k \leq \pi$. It suffices to show $\theta_k$ is not a rational multiple of $\pi$. Recall that $q = e^{\frac{2\pi i}{k+2}}$. We have,
\begin{align}
     2\cos \theta_k = \operatorname{tr}(A) &=-\frac{((q-1) q+1) \left(q^2+1\right)}{q^2} \\
     &= -2 - (q^2 + \frac{1}{q^2}) + (q + \frac{1}{q})\\
     &= -2 - 2\cos \frac{4 \pi}{k+2} + 2\cos \frac{2 \pi}{k+2}.
\end{align}
That is,
\begin{equation}\label{eqn:matrixA_trace_identity}
     \cos{\frac{4\pi}{k+2}}-\cos{\frac{2\pi}{k+2}}+\cos{\theta_k}= -1.
\end{equation}
We wish to show the above identity is not equivalent to any of the identities in Theorem \ref{thm:cosin_sum}, and hence $\theta_k$ cannot be a rational multiple of $\pi$. But we need to first verify the assumptions in that theorem. 

We make four statements which can be verified by checking small values of $k$ and applying Theorem \ref{thm:cosin_sum}, noting that when $k > 6$,   $\frac{2\pi}{k+2}$ and $\frac{4\pi}{k+2}$  lie strictly between 0 and $\frac{\pi}{2}$. Assume $k \geq 3$.
\begin{enumerate}[label=\Alph*)., leftmargin = *]
    \item 
The only $k$ for which $\cos{\frac{2\pi}{k+2}}$ is rational is $k=4$,
\begin{align}
    \cos{\frac{2\pi}{4+2}} = \frac{1}{2}.
\end{align}
    \item 
The only $k'$s for which $\cos{\frac{4\pi}{k+2}}$ is rational are $k=4, 6, 10$,
\begin{align}
    \cos{\frac{4\pi}{4+2}} = -\frac{1}{2}, \quad \cos{\frac{4\pi}{6+2}} = 0, \quad \cos{\frac{4\pi}{10+2}} = \frac{1}{2}.
\end{align}
    \item 

The only $k'$s for which neither $\cos{\frac{2\pi}{k+2}}$ nor $\cos{\frac{4\pi}{k+2}}$ is rational but certain non-trivial rational combinations of them is rational are $k =3, 8$,
\begin{align}
-\cos{\frac{2\pi}{3+2}} - \cos{\frac{4\pi}{3+2}} = \frac{1}{2}, \quad\quad
    \cos{\frac{2\pi}{8+2}} - \cos{\frac{4\pi}{8+2}} = \frac{1}{2}.
\end{align}

    \item 
Furthermore, the only $k'$s for which $\cos{\theta_k}$ is rational are $k = 4, 8$,
\begin{align}
    \cos \theta_4 = 0, \quad \cos \theta_8 = -\frac{1}{2}.
\end{align}
\end{enumerate}

Statement $D)$ implies that for $k = 4, 8$, $\theta_k$ is a rational multiple of $\pi$. For all other $k'$s, $\theta_k$ is not a rational multiple of $\pi$ and in particular $\theta_k \neq 0, \frac{\pi}{2}, \pi$.

The above statements also imply for $k \geq 7$, $k \neq 8, 10$, none of $\cos{\frac{4\pi}{k+2}}$, $\cos{\frac{2\pi}{k+2}}$, or $\cos{\theta_k}$ is rational.  Furthermore, for those values of $k$, for any two elements from $\{\cos{\frac{4\pi}{k+2}}, \cos{\frac{2\pi}{k+2}}, \cos{\theta_k} \}$, they cannot have a non-trivial rational combination which is rational. The above claim for the pair $\{\cos{\frac{4\pi}{k+2}}, \cos{\frac{4\pi}{k+2}}\}$ is clear from Statement $C)$. For the other two pairs, say, $\{\cos{\frac{4\pi}{k+2}}, \cos{\theta_k}\}$, if they do not satisfy the claim, then one can substitute $\cos{\theta_k}$ with an expression of $\cos{\frac{4\pi}{k+2}}$ in Equation \ref{eqn:matrixA_trace_identity}, yielding a rational combination of $\cos{\frac{4\pi}{k+2}}$ and $\cos{\frac{2\pi}{k+2}}$ which is rational. That is a contradiction.

Then for $k \geq 7$, $k \neq 8, 10$, if $\theta_k$ is a rational multiple of $\pi$ strictly between 0 and $\frac{\pi}{2}$, then Equation \ref{eqn:matrixA_trace_identity} is an identity concerning a rational combination of the cosine of three distinct angles satisfying the conditions in Theorem \ref{thm:cosin_sum}. However, this identity is not equivalent to any of those in that theorem, a contradiction. 

If $\theta_k$ is a rational multiple of $\pi$ strictly between $\frac{\pi}{2}$ and $\pi$, then Equation \ref{eqn:matrixA_trace_identity} can be written as,
\begin{equation}
     \cos{\frac{4\pi}{k+2}}-\cos{\frac{2\pi}{k+2}}-\cos(\pi -\theta_k)= -1.
\end{equation}
Applying the same argument to the new equation lead to a similar contradiction. This shows that for $k \geq 7$, $k \neq 8, 10$, $\theta_k$ is not a  rational multiple of $\pi$.

The remaining cases to check are $k = 3, 5, 6, 10$. The corresponding identity of Equation \ref{eqn:matrixA_trace_identity} for these values of $k$ can be simplified below.
\begin{align}
    \cos{\theta_3} = \frac{\sqrt{5}-2}{2}, \ 
    \cos{\theta_5} = \cos{\frac{3\pi}{7}}+\cos{\frac{2\pi}{7}} -1, \ 
    \cos{\theta_6} = \frac{\sqrt{2}-2}{2}, \ 
    \cos{\theta_{10}} = \frac{\sqrt{3}-3}{2}.
\end{align}
It can be checked directly by a computer program or by applying Theorem \ref{thm:cosin_sum} that the above $\theta_k'$s are not rational multiple of $pi$. This completes the proof that the eigenvalues of $A$ are of infinite order for $k \geq 3$, $k \neq 4, 8$.

\vspace{0.2cm}
\noindent\textbf{Proving $\mathbf{B}$ has infinite order.} 

Denote the eigenvalues of $B$ by $e^{\pm \theta_k i}$, $0 \leq \theta_k \leq \pi$. 
\begin{align}
     2\cos \theta_k = \operatorname{tr}(B) &= \frac{\left(q^2+1\right) \left((q-1) q \left(q^2+1\right)+1\right)}{q^3} \\
     &= -2 + (q^3 + \frac{1}{q^3}) - (q^2 + \frac{1}{q^2}) + 2(q + \frac{1}{q})\\
     &= -2 + 2\cos \frac{6 \pi}{k+2} - 2\cos \frac{4 \pi}{k+2} + 4\cos \frac{2 \pi}{k+2}.
\end{align}
That is,
\begin{equation}
     2\cos \frac{2 \pi}{k+2}  - \cos \frac{4 \pi}{k+2} + \cos \frac{6 \pi}{k+2} - \cos{\theta_k}= 1.
\end{equation}
The rest of the proof is completely similar to the case of the matrix $A$ by repeatedly applying Theorem \ref{thm:cosin_sum}. We leave the details as an exercise for curious readers.

\vspace{0.2cm}
\noindent\textbf{Proving $\mathbf{W \neq I}$.}

Note that $W \neq I$ if and only if $\operatorname{Tr}(W) \neq 2$. Assume $\operatorname{Tr}(W) = 2$. Then,
\begin{align}
     2 &= -\frac{((q-1) q+1) \left(q^4-2 q^3-2 q+1\right)}{q^3} \\
       &= 4 - (q^3 + \frac{1}{q^3}) + 3 (q^2 + \frac{1}{q^2}) - 3(q + \frac{1}{q})\\
     &= 4 - 2\cos \frac{6 \pi}{k+2} + 6\cos \frac{4 \pi}{k+2} - 6\cos \frac{2 \pi}{k+2}.
\end{align}
That is,
\begin{equation}
    \cos \frac{6 \pi}{k+2} - 3 \cos \frac{4 \pi}{k+2} + 3 \cos \frac{2 \pi}{k+2} = 1.
\end{equation}
That the above identity is fake can be checked directly for small values $k \leq 10$ and by Theorem \ref{thm:cosin_sum} for $k \geq 11$.

    \end{proof}
\end{theorem}

\appendix

\section{Anyon models}\label{sec:anyon_model_detail}
Mathematically, an anyon model is characterized by the structure of a unitary modular tensor category (MTC). To avoid abstract categorical language, we describe an MTC with a set of concrete data. The content and figures contained in this section are adapted from the second named author's lecture notes \cite{cui2018lecture}.

\vspace{0.1cm}
\noindent\textbf{Label set.} Associated with each anyon model is a finite set 
\begin{equation}
L = \{a,b,c, \cdots\}
\end{equation}
consisting of all the possible anyon types in a topological phase. The ground state is considered as a special trivial anyon type, and is usually denoted by $\mathbf{1} \in L$. For each anyon type $x \in L$, there exists $\bar{x} \in L$ corresponding to the anti-particle (i.e., the dual anyon type) of $x$. We require that $\bar{\mathbf{1}} = \mathbf{1}$ and $\bar{\bar{x}} = x$. 

\vspace{0.1cm}
\noindent\textbf{Fusion rule.} For $a,b \in L$, fusing $a$ and $b$ produces different possible anyon types. Formally, it is written as,
\begin{equation}
    a \otimes b = \bigoplus_{c \in L}\ N_{ab}^c\ c,
\end{equation}
where $N_{ab}^c$ denotes the number of different channels of fusing $a$ and $b$ to result in the output $c$. If there is no way to obtain  $c$ from the fusion, then $N_{ab}^c = 0$. One can also view the equality in the above equation from an alternative perspective. Namely, the composite type of $a$ and $b$ is a superposition of all possible anyon types with each type $c$ appearing in $N_{ab}^c$ copies.
If $N_{ab}^c > 0$, we say $c$ is a \emph{total type} or \emph{total charge} of $a$ and $b$, and call the triple $(a,b;c)$ \emph{admissible}. The collection of the integers $\{N_{ab}^c\ |\ a,b,c \in L\}$ is called the fusion rule. The fusion rule should satisfy the following requirements.
\begin{enumerate}[label=\Alph*)., leftmargin = *]
    \item The fusion rule is commutative, i.e., $a \otimes b = b \otimes a$, implying
     \begin{align}
N_{ab}^c = N_{ba}^c, \quad\forall a,b,c.
\end{align} 

\item The dual of $a \otimes b$ as a composite equals $\bar{b} \otimes \bar{a}$, implying
\begin{align}
N_{ab}^c = N_{\bar{b}\bar{a}}^{\bar{c}}, \quad\forall a,b,c.
\end{align}

\item $\mathbf{1} \otimes a = a$, implying
\begin{align}
N_{\mathbf{1} a}^b = \delta_{a,b}, \quad\forall a,b.
\end{align}

\item $\mathbf{1}$ is a total type of $a$ and $b$ if and only if $a = \bar{b}$, implying
\begin{align}
N_{ab}^{\mathbf{1}} = \delta_{a, \bar{b}}, \quad\forall a,b.
\end{align}

\item The fusion rule is associative, i.e., $(a \otimes b) \otimes c = a \otimes (b \otimes c)$, implying
\begin{align}
\sum\limits_{p \in L} N_{ab}^p N_{pc}^d = \sum\limits_{q \in L} N_{aq}^d N_{bc}^q, \quad\forall a,b,c,d,
\end{align} 
\end{enumerate}

For simplicity, in the following discussions \textbf{we will assume $N_{ab}^c$ is either $0$ or $1$}, i.e., the anyon model is multiplicity-free. This already covers a large family of anyon models including the ones considered in the current paper. 

\vspace{0.1cm}
\noindent\textbf{State space.} For anyon types $c, a_1, \cdots, a_n$, denote by $V_{c}^{a_1a_2\cdots a_n}$ the space of states representing $n$ anyons $a_1, \cdots, a_n$ with total charge $c$. The dimension and bases of these spaces are described inductively as follows. 

\begin{figure}
\centering
\begin{tikzpicture}[thick]
\begin{scope}
\draw (0,1) -- (0,-1) node[right]{$c$};
\end{scope}

\begin{scope}[xshift = 5cm]
\draw (0,0) -- (0,-1) node[right]{$c$};
\draw (0,0) -- (-1,1) node[left]{$a_1$};
\draw (0,0) -- (1,1) node[right]{$a_2$};
\end{scope}
\end{tikzpicture}
\caption{(Left) A canonical basis of $V_{c}^c$; (Right) A (non-canonical) basis of $V_{c}^{a_1a_2}$.}\label{fig:basis_one_and_two_anyons}
\end{figure}
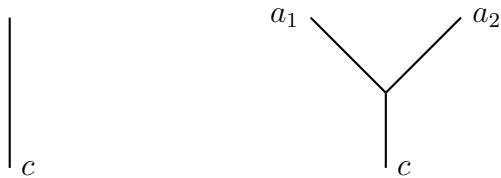

For $n = 1$, $V_{c}^{a_1} = 0$ if $c \neq a_1$, and $V_{c}^{c}$ is $1$-dimensional with a canonical basis denoted by the left diagram of Figure \ref{fig:basis_one_and_two_anyons}. One can think of the diagram representing the state obtained by `doing nothing' to an existent anyon $c$.

For $n = 2$, $V_{c}^{a_1a_2} = 0$ if $N_{a_1a_2}^c = 0$, and $V_{c}^{a_1a_2}$ is $1$-dimensional otherwise, with a non-canonical basis denoted by the right diagram in Figure \ref{fig:basis_one_and_two_anyons}. One can then think of the diagram representing the state obtained from the process of splitting $c$ into the pair $a$ and $b$, where and throughout the context, the time direction for physical processes is assumed to be from bottom to top. This choice of basis is not canonical as one can multiply an arbitrary phase to it. We will use the diagrams in Figure \ref{fig:basis_one_and_two_anyons} as building blocks to describe bases of multi-anyon spaces.

\begin{figure}
\begin{tikzpicture}
\draw[thick] (0.5,-0.5) -- (-1,1)node[above]{$a_1$};
\draw[thick] (0,1)node[above]{$a_2$} -- (-0.5,0.5)node[below]{$b_1$};
\draw[thick] (0,0)node[below]{$b_2$} -- (1,1)node[above]{$a_3$};
\draw[thick] (2,1)node[above]{$a_4$} -- (0.5,-0.5);
\draw (3,1) node{$\cdots$};
\draw[thick] (4,1)node[above]{$a_n$} -- (1.5,-1.5);

\draw[thick] (0.5,-0.5) -- (2,-2)node[below]{$c$};
\fill[white] (1,-1) circle[radius = 0.3cm];
\draw (1,-1) node[rotate = -45]{$\cdots$};
\end{tikzpicture}
\caption{A basis of $V_{c}^{a_1a_2\cdots a_n}$ corresponding to a binary tree}\label{fig:basis_n_anyons}
\end{figure}
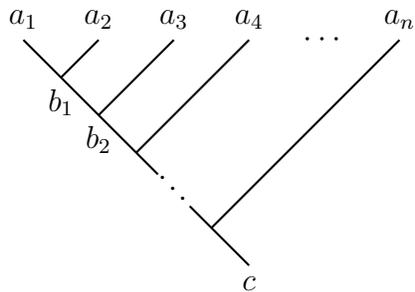

For $n \geq 3$, take an upward-growing binary tree with one root at the bottom and $n$ leaves at the top. See Figure \ref{fig:basis_n_anyons} for an illustration. It is to be understood that the tree is constructed using the two diagrams in Figure \ref{fig:basis_one_and_two_anyons}. Label the root by $c$ and the leaves, from left to right, by $a_1, a_2, \cdots, a_n$. Now label each internal edge $e$ by an anyon type $b_e$ such that at each fork, the relevant labels are admissible. Then the binary tree with all possible labels $\{b_e\}$ of internal edges form a basis of  $V_{c}^{a_1a_2\cdots a_n}$. For each labeled binary tree, one can similarly interpret the state it represents as a splitting process. For example, the state represented by the tree in Figure \ref{fig:basis_n_anyons} is obtained by splitting $c$ into $b_{n-2}$ and $a_n$, followed by splitting $b_{n-2}$ into $b_{n-3}$ and $a_{n-1}$, $\cdots\cdots$, followed by splitting $b_1$ into $b_0 = a_1$ and $a_2$. Such a basis is called a \emph{splitting-tree basis}. 

For the case of $n=3$, there are exactly two such binary trees as shown on both sides of Equation \ref{eqn:F_symbol_def}, each of which provides a basis of $V_d^{abc}$. Each tree has one internal edge. The basis corresponding to the tree on the left side of the equation consists of all possible labelings $m$ of the internal edge so that $(a,b;m)$ and $(m,c;d)$ are both admissible. Similarly, the basis for the tree on the right side consists of labelings $n$ of the internal edge so that $(b,c;n)$ and $(a,n;d)$ are both admissible. Denote the matrix change between the two bases by $F_{d}^{abc}$. More explicitly,
\begin{equation}\label{eqn:F_symbol_def}
\begin{aligned}
\begin{tikzpicture}
\begin{scope}
\draw[thick] (0.5,-0.5)node[below]{$d$} --(0,0) -- (-1,1)node[above]{$a$};
\draw[thick] (0,0) -- (1,1)node[above]{$c$};
\draw[thick] (0,1)node[above]{$b$} -- (-0.5,0.5)node[below]{$m$};
\end{scope}

\draw (2,0.25) node{$= \, \sum\limits_{n} \, F^{abc}_{d;nm}$};

\begin{scope}[xshift = 4cm]
 \draw[thick] (0,0) -- (-1,1)node[above]{$a$};
 \draw[thick] (-0.5,-0.5)node[below]{$d$} --(0,0) -- (1,1)node[above]{$c$};
 \draw[thick] (0,1)node[above]{$b$} -- (0.5,0.5)node[below]{$n$};
\end{scope}
\end{tikzpicture}
\end{aligned}
\end{equation}
where $F^{abc}_{d;nm}$ is the $(n,m)$-entry of $F_{d}^{abc}$, and the sum is over all labelings $n$ as described above. Note that, here the anyon types $n$ and $m$ are used as the indices of the entries of $F_{d}^{abc}$.  We call $F_{d}^{abc}$ an $F$-matrix, its entries $F$-symbols or $6j$-symbols, and the identity in Equation \ref{eqn:F_symbol_def} an $F$-move. 

\begin{figure}
\begin{tikzpicture}[thick]
\begin{scope}[xshift = 0cm, yshift = 0cm]
\draw (0,0) -- (-1.5,1.5)node[above]{$a$};
\draw (-1, 1)node[below]{$m$} -- (-0.5,1.5)node[above]{$b$};
\draw (-0.5, 0.5)node[below]{$n$} -- (0.5,1.5)node[above]{$c$};
\draw (0,0) -- (1.5,1.5)node[above]{$d$};
\draw (0,0)-- (0.5,-0.5)node[below]{$e$};
\draw (0.5,-1.2) node{$\circled{1}$};

\draw (-2,0) node(topL){};
\draw (2,0) node(topR){};
\end{scope}
\begin{scope}[xshift = -5.52cm,yshift = -4cm]
\draw (0,0) -- (-1.5,1.5)node[above]{$a$};
\draw (-1, 1)node[below]{$m$} -- (-0.5,1.5)node[above]{$b$};
\draw (1, 1)node[below]{$z$} -- (0.5,1.5)node[above]{$c$};
\draw (0,0) -- (1.5,1.5)node[above]{$d$};
\draw (0,0)-- (0,-0.5)node[below]{$e$};
\draw (0,-1.2) node{$\circled{2}$};

\draw (0.5, 2.5)node(middle1T){};
\draw (1, -0.5)node(middle1B){};
\end{scope}
\begin{scope}[xshift = -3.5cm, yshift = -9cm]
\draw (0,0) -- (-1.5,1.5)node[above]{$a$};
\draw (0.5, 0.5)node[below]{$y$} -- (-0.5,1.5)node[above]{$b$};
\draw (1, 1)node[below]{$z$} -- (0.5,1.5)node[above]{$c$};
\draw (0,0) -- (1.5,1.5)node[above]{$d$};
\draw (0,0)-- (-0.5,-0.5)node[below]{$e$};
\draw (-0.5,-1.2) node{$\circled{3}$};

\draw (0.5, 2.5)node(bottom1T){};
\draw (1.5, 0.5)node(bottom1R){};
\end{scope}
\begin{scope}[xshift = 5.52cm, yshift = -4cm]
\draw (0,0) -- (-1.5,1.5)node[above]{$a$};
\draw (0, 1)node[below]{$x$} -- (-0.5,1.5)node[above]{$b$};
\draw (-0.5, 0.5)node[below]{$n$} -- (0.5,1.5)node[above]{$c$};
\draw (0,0) -- (1.5,1.5)node[above]{$d$};
\draw (0,0)-- (0.5,-0.5)node[below]{$e$};
\draw (0.5,-1.2) node{$\circled{5}$};

\draw (-0.5, 2.5)node(middle2T){};
\draw (-0.5, -0.5)node(middle2B){};
\end{scope}
\begin{scope}[xshift = 3.5cm, yshift = -9cm]
\draw (0,0) -- (-1.5,1.5)node[above]{$a$};
\draw (0.5, 0.5)node[below]{$y$} -- (-0.5,1.5)node[above]{$b$};
\draw (0, 1)node[below]{$x$} -- (0.5,1.5)node[above]{$c$};
\draw (0,0) -- (1.5,1.5)node[above]{$d$};
\draw (0,0)-- (-0.5,-0.5)node[below]{$e$};
\draw (-0.5,-1.2) node{$\circled{4}$};

\draw (-0.5, 2.5)node(bottom2T){};
\draw (-1.5, 0.5)node(bottom2L){};
\end{scope}
\draw[->, >=stealth,red] (topL) -- (middle1T); 
\draw[->, >=stealth,red] (topR) -- (middle2T); 
\draw[->, >=stealth,red] (middle1B) -- (bottom1T);
\draw[->, >=stealth,red] (middle2B) -- (bottom2T);
\draw[->, >=stealth,red] (bottom2L) -- (bottom1R);
\end{tikzpicture}
\caption{The five splitting-tree bases of $V_{e}^{abcd}$ and the $F$-moves connecting them.}\label{fig:F_move_pentagon}
\end{figure}
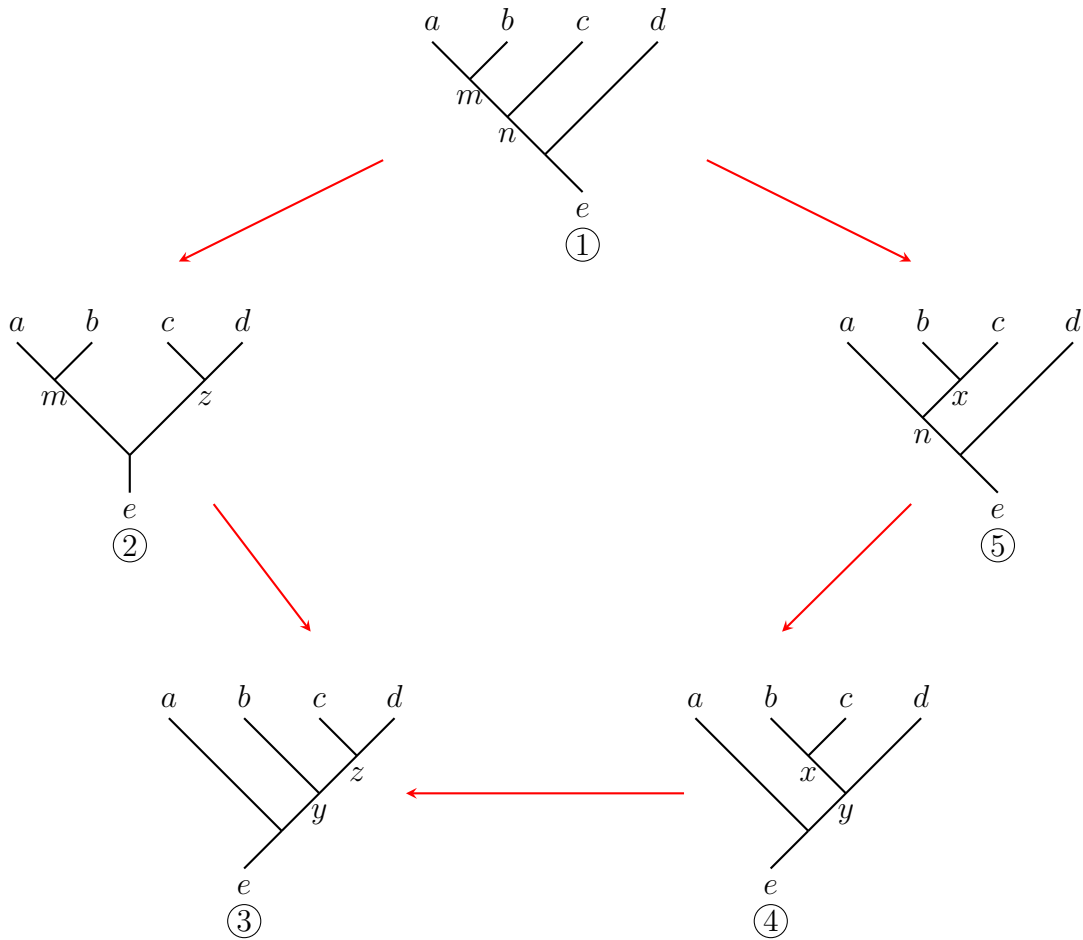

Using the $F$-move or its inverse, we can relate any two splitting-tree bases of $V_{c}^{a_1a_2\cdots a_n}$. For $V_{e}^{abcd}$, there are exactly five splitting-tree bases, and Figure \ref{fig:F_move_pentagon} shows the $F$-moves connecting them. In particular, starting from the basis labeled by $\circled{1}$, there are two ways of performing $F$-moves to obtain the basis labeled by $\circled{3}$, namely, either via the path $\circled{1} \to \circled{2} \to \circled{3}$ or via the path $\circled{1} \to \circled{5} \to \circled{4} \to \circled{3}$. Since both ways induce a basis change between $\circled{1}$ and $\circled{3}$, this introduces some constraints on the $F$-symbols, namely,
\begin{align}
\label{equ:pentagon}
F^{mcd}_{e;zn}F^{abz}_{e;ym} = \sum\limits_{x \in L} F^{abc}_{n;xm}F^{axd}_{e;yn}F^{bcd}_{y;zx}, \quad\forall a,b,c,d,e,m,n,y,z.
\end{align} 
Equation \ref{equ:pentagon} is known as the Pentagon Equations. It is a non-trivial fact that the Pentagon Equations guarantee that the change between splitting-tree bases via $F$-moves for an arbitrary state space is consistent.

\vspace{0.1cm}
\noindent\textbf{Braiding.} The process of swapping positions of anyons is called a braiding. Since anyons  live in 2-dimensional space, a counterclockwise braiding has a different world-line from that of a clockwise braiding. The world-line of a sequence of braidings of multi-anyons is a braid diagram, and hence the naming of the process. A braiding induces a unitary transformation on the state space. Consider two anyons $a$ and $b$ with total type $c$. A counterclockwise braiding of $a$ and $b$ maps a state in $V_{c}^{ab}$ to one  in $V_{c}^{ba}$. Since both spaces have dimension one, there exists a phase $R^{ba}_c$ such that the following equality holds,
\begin{equation}
\begin{aligned}
\begin{tikzpicture}[thick]
\begin{scope}
\draw (0,0) -- (-0.5,0.5)node[left]{$a$};
\draw (0,0) -- (0.5,0.5)node[right]{$b$};
\draw (0,0) -- (0,-0.5)node[below]{$c$};
\braid  (br) at (-0.5, 2) s_1^{-1};
\draw (br-1-s) node[left]{$b$};
\draw (br-2-s) node[right]{$a$};
\end{scope}

\draw (1.5,0.5) node{$=$};
\draw (2.5,0.5) node{$R^{ba}_{c}$};

\begin{scope}[xshift = 3.5cm]
\draw (0,0) -- (-0.5,0.5) -- (-0.5, 2)node[left]{$b$};
\draw (0,0) -- (0.5,0.5) -- (0.5, 2)node[right]{$a$};
\draw (0,0) -- (0,-0.5)node[below]{$c$};
\end{scope}
\end{tikzpicture}
\end{aligned}
\end{equation}
The above equality is called an $R$-move, and $\{R^{ab}_c\}$ is called an $R$-symbol.  Since a counterclockwise braiding followed by a clockwise braiding is equivalent to the identity process, we have,
\begin{equation}
\begin{aligned}
\begin{tikzpicture}[thick]
\begin{scope}
\draw (0,0) -- (-0.5,0.5)node[left]{$a$};
\draw (0,0) -- (0.5,0.5)node[right]{$b$};
\draw (0,0) -- (0,-0.5)node[below]{$c$};
\braid (br) at (-0.5, 2) s_1;
\draw (br-1-s) node[left]{$b$};
\draw (br-2-s) node[right]{$a$};
\end{scope}

\draw (1.5,0.5) node{$=$};
\draw (2.5,0.5) node{$(R^{ab}_{c})^{-1}$};

\begin{scope}[xshift = 4cm]
\draw (0,0) -- (-0.5,0.5) -- (-0.5, 2)node[left]{$b$};
\draw (0,0) -- (0.5,0.5) -- (0.5, 2)node[right]{$a$};
\draw (0,0) -- (0,-0.5)node[below]{$c$};
\end{scope}
\end{tikzpicture}
\end{aligned}
\end{equation}

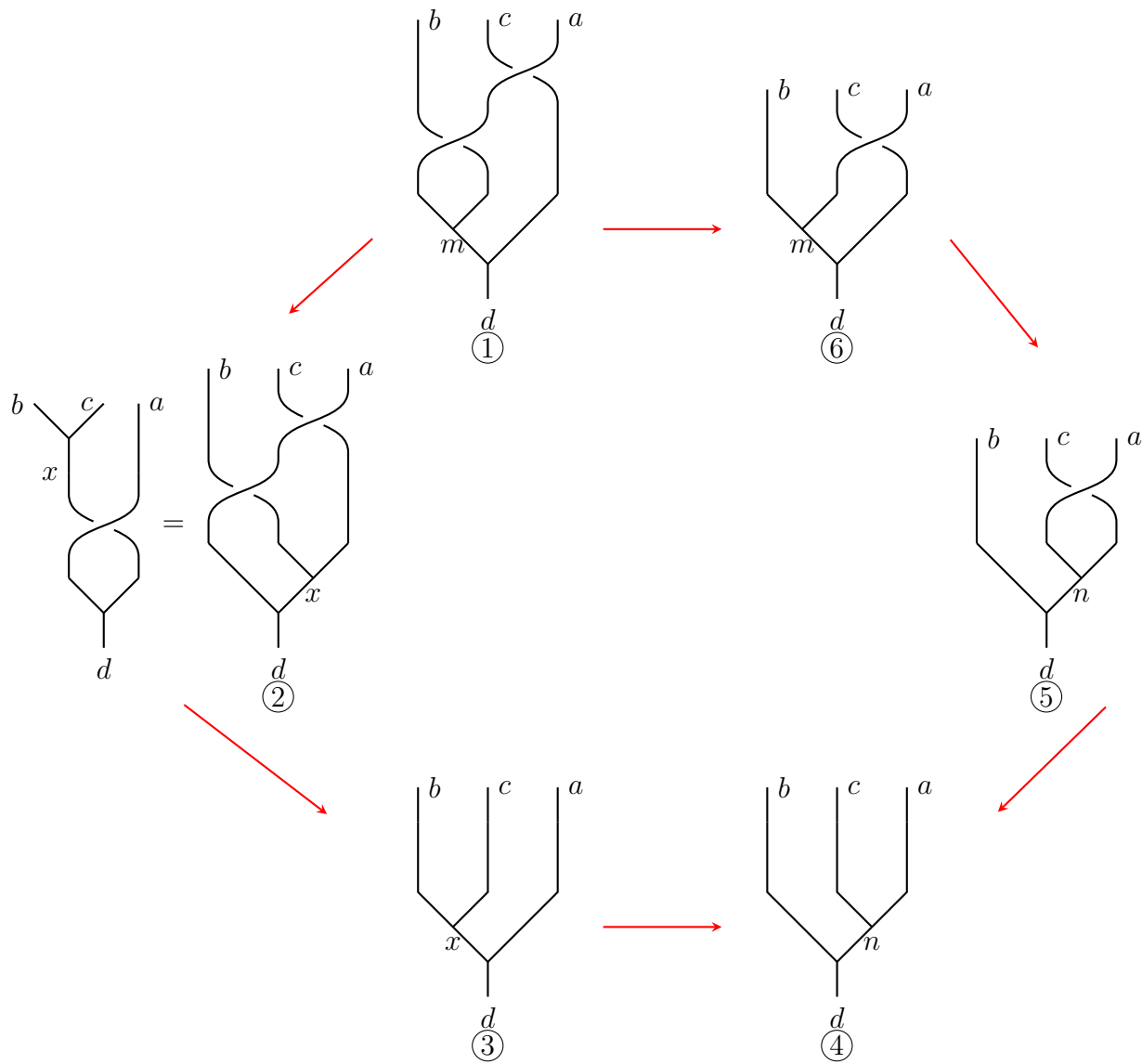
\begin{figure}
\begin{tikzpicture}[thick]
\begin{scope}[xshift = 0cm, yshift = 0cm]
\draw (0,-0.5)node[below]{$d$} --(0,0) -- (-1,1);
\draw (0,0) -- (1,1);
\draw (0,1) -- (-0.5,0.5)node[below]{$m$};
\braid (br) at (-1, 3.5) s_2^{-1}s_1^{-1};
\draw (br-1-s)node[right]{$b$};
\draw (br-2-s)node[right]{$c$};
\draw (br-3-s)node[right]{$a$};

\draw (0,-1.2) node{$\circled{1}$};
\draw (-1.5,0.5) node(top1L){};
\draw (1.5,0.5) node(top1R){};
\end{scope}

\begin{scope}[xshift = 5cm, yshift = 0cm]
\draw (0,-0.5)node[below]{$d$} --(0,0) -- (-1,1);
\draw (0,0) -- (1,1);
\draw (0,1) -- (-0.5,0.5)node[below]{$m$};
\braid[number of strands = 3] (br) at (-1, 2.5) s_2^{-1};
\draw (br-1-s)node[right]{$b$};
\draw (br-2-s)node[right]{$c$};
\draw (br-3-s)node[right]{$a$};

\draw (0,-1.2) node{$\circled{6}$};
\draw (-1.5,0.5) node(top2L){};
\draw (1.5,0.5) node(top2R){};
\end{scope}

\begin{scope}[xshift = 8cm, yshift = -5cm]
\draw (0,-0.5)node[below]{$d$} --(0,0) -- (-1,1);
\draw (0,0) -- (1,1);
\draw (0,1) -- (0.5,0.5)node[below]{$n$};
\braid[number of strands = 3] (br) at (-1, 2.5) s_2^{-1};
\draw (br-1-s)node[right]{$b$};
\draw (br-2-s)node[right]{$c$};
\draw (br-3-s)node[right]{$a$};

\draw (0,-1.2) node{$\circled{5}$};
\draw (br-2-s)node[above, yshift = 1cm](middle2T){};
\draw (1, -1.2) node(middle2B) {};
\end{scope}

\begin{scope}[xshift = 5cm, yshift = -10cm]
\draw (0,-0.5)node[below]{$d$} --(0,0) -- (-1,1) -- (-1,2);
\draw (0,0) -- (1,1) -- (1,2);
\draw (0,2) -- (0,1) -- (0.5,0.5)node[below]{$n$};
\braid[number of strands = 3] (br) at (-1, 2.5) id;
\draw (br-1-s)node[right]{$b$};
\draw (br-2-s)node[right]{$c$};
\draw (br-3-s)node[right]{$a$};

\draw (0,-1.2) node{$\circled{4}$};
\draw (br-3-s)node[right, xshift = 1cm, yshift = -0.5cm](bottom2T){};
\draw (-1.5, 0.5) node(bottom2L) {};
\end{scope}

\begin{scope}[xshift = 0cm, yshift = -10cm]
\draw (0,-0.5)node[below]{$d$} --(0,0) -- (-1,1) -- (-1,2);
\draw (0,0) -- (1,1) -- (1,2);
\draw (0,2) -- (0,1) -- (-0.5,0.5)node[below]{$x$};
\braid[number of strands = 3] (br) at (-1, 2.5) id;
\draw (br-1-s)node[right]{$b$};
\draw (br-2-s)node[right]{$c$};
\draw (br-3-s)node[right]{$a$};

\draw (0,-1.2) node{$\circled{3}$};
\draw (br-1-s)node[left, yshift = -0.5cm, xshift = -1cm](bottom1T){};
\draw (1.5, 0.5) node(bottom1R) {};
\end{scope}

\begin{scope}[xshift = -3cm, yshift = -5cm]
\draw (0,-0.5)node[below]{$d$} --(0,0) -- (-1,1);
\draw (0,0) -- (1,1);
\draw (0,1) -- (0.5,0.5)node[below]{$x$};
\braid (br) at (-1, 3.5) s_2^{-1}s_1^{-1};
\draw (br-1-s)node[right]{$b$};
\draw (br-2-s)node[right]{$c$};
\draw (br-3-s)node[right]{$a$};
\draw (-1.5, 1.25) node{$=$};
\begin{scope}[xshift = -2.5cm]
\draw (0,-0.5)node[below]{$d$} --(0,0) -- (-0.5,0.5);
\draw (0,0) -- (0.5,0.5);
\braid (br2) at (-0.5, 2) s_1^{-1};
\draw (-0.5,2)node[left]{$x$} --(-0.5,2.5) -- (-1,3)node[left]{$b$};
\draw (-0.5,2.5) -- (0,3)node[left]{$c$};
\draw (0.5,2) -- (0.5,3) node[right]{$a$};
\end{scope}

\draw (0,-1.2) node{$\circled{2}$};
\draw (br-2-s) node[above, yshift = 0.5cm](middle1T){};
\draw (-1.5, -1.2) node(middle1B){};
\end{scope}

\draw[->, >=stealth, red] (top1L) -- (middle1T);
\draw[->, >=stealth, red] (middle1B) -- (bottom1T);
\draw[->, >=stealth, red] (bottom1R) -- (bottom2L);
\draw[->, >=stealth, red] (top1R) -- (top2L);
\draw[->, >=stealth, red] (top2R) -- (middle2T);
\draw[->, >=stealth, red] (middle2B) -- (bottom2T);
\end{tikzpicture}
\caption{Consistency conditions for braidings on $V_{d}^{abc}$.}\label{fig:braiding_hexagon}
\end{figure}

Consider the space $V_{d}^{abc}$, and braid $a$ with $b$ and $c$. Each node in Figure \ref{fig:braiding_hexagon} represents a basis of $V_{d}^{bca}$, and performing an $F$-move or $R$-move change from one basis to another. To change from basis $\circled{1}$ to basis $\circled{3}$, one can follow either  the path $\circled{1} \to \circled{2} \to \circled{3}$ or the path $\circled{1} \to \circled{6} \to \circled{5} \to \circled{4} \to \circled{3}$. Consequently, we obtain the Hexagon Equation,
\begin{align}
\label{equ:hexagon1}
R^{ba}_{m}F^{bac}_{d;nm}R^{ca}_{n} = \sum\limits_{x \in L} F^{abc}_{d;xm}R^{xa}_{d}F^{bca}_{d;nx}.
\end{align}
By replacing the counterclockwise braidings in Figure \ref{fig:braiding_hexagon} with clockwise braidings, we obtain another Hexagon Equation,
\begin{align}
\label{equ:hexagon2}
(R^{ab}_{m})^{-1}F^{bac}_{d;nm}(R^{ac}_{n})^{-1} = \sum\limits_{x \in L} F^{abc}_{d;xm}(R^{ax}_{d})^{-1}F^{bca}_{d;nx}.
\end{align}

\vspace{0.1cm}
\noindent\textbf{Topological spin.} Each anyon type $a$ has an (intrinsic) topological spin $\theta_a$ which is always a root of unity. The type $a$ is said to be bosonic if $\theta_a = 1$, fermionic if $\theta_a = -1$, and semionic if $\theta_a = i$. The topological spins are required to satisfy the following conditions.
\begin{enumerate}[label=\Alph*)., leftmargin = *]
\item The trivial anyon is bosonic,
\begin{equation}
    \theta_{\mathbf{1}} = 1.
\end{equation}
\item An anyon and its dual have equal topological spin,
\begin{equation}
    \theta_{a} = \theta_{\bar{a}}, \quad\forall a \in L.
\end{equation}
\item Whenever $c$ is a total type of $a$ and $b$, we have
\begin{equation}
    \theta_c \theta_a^{-1} \theta_b^{-1} = R^{ab}_cR^{ba}_c.
\end{equation}
\end{enumerate}

\vspace{0.1cm}
\noindent\textbf{Quantum dimension.} For each anyon type $a$, define an $|L| \times |L|$ matrix $N_a$ whose $(b,c)$-entry is $N_{ab}^c = N_{ba}^c$. Hence the entries of $N_{a}$ are non-negative integers. By the Perron-Frobenius theorem, $N_a$ has an eigenvalue $\dim (a)$, called the Frobenius-Perron dimension of $a$, which is greater than or equal to, in absolute value, any other eigenvalues. In the anyon model, we also call $\dim (a)$ the quantum dimension of $a$. To get a sense of what $\dim (a)$ measures, consider the dimension of the space of $n$ type-$a$ anyons with total type $\mathbf{1}$. We use the splitting-tree basis in Figure \ref{fig:basis_n_anyons} with $a = a_1 = \cdots = a_n, \ \mathbf{1} = c$ to compute the dimension for large $n$,
\begin{align}
\sum\limits_{b_1, \cdots, b_{n-2}} N_{aa}^{b_1}N_{b_1a}^{b_2} \cdots N_{b_{n-3}a}^{b_{n-2}}N_{b_{n-2}a}^{\mathbf{1}} &= \sum\limits_{b_1, \cdots, b_{n-3}} N_{aa}^{b_1}N_{b_1a}^{b_2} \cdots N_{b_{n-3}a}^{\bar{a}} \\
&= \left((N_a)^{n-2}\right)_{a\bar{a}} \quad \overset{n \rightarrow \infty}{\sim} \quad \dim (a)^{n-2}. 
\end{align}
Thus, $\dim (a)$ measures the asymptotic size of the space of $n$ type-$a$ anyons. Apparently, $\dim (a) \geq 1$. An anyon $a$ is called \emph{Abelian} if $\dim(a) = 1$, and \emph{non-Abelian} otherwise.

\vspace{0.1cm}
\noindent\textbf{$S$-matrix.} Define the $|L| \times |L|$ modular $S$-matrix with entries,
\begin{equation}
    S_{ab}:= \theta_a^{-1}\theta_b^{-1} \sum_{c \in L} N_{\bar{a}b}^c \ \theta_c \dim(c).
\end{equation}
The $S$-matrix is required to be invertible.

To summarize, an anyon model or a unitary MTC is described by a label set, fusion rule, $F$-symbols, $R$-symbols, and topological spins, from which one can derive quantum dimensions and the $S$-matrix. These data should satisfy various compatibility conditions as listed in this section.  

\section{Mathematica code}\label{sec:Mathematica}
Below we list the Mathematica code to compute some of the matrices used in Section \ref{subsec:1_qubit_model_SU2} and \ref{subsec:main_result} including $F$, $\tilde{R}$, $A$, $B$, and $W$.

\begin{lstlisting}[
                   language = Mathematica,
                   ]

Clear[q];
QuantumInteger[n_] := (q^(n/2) - q^(-n/2))/(q^(1/2) - q^(-1/2));
F = FullSimplify[{{-1, 
      Sqrt[QuantumInteger[3]]}, {Sqrt[QuantumInteger[3]], 1}}/
    QuantumInteger[2]];
Rtilde = DiagonalMatrix[{q^(-1/2), -q^(1/2)}]*I;
A = Simplify[
   MatrixPower[Rtilde, 2] . F . MatrixPower[Rtilde, 4] . F];
B = Simplify[
   MatrixPower[Rtilde, 2] . F . MatrixPower[Rtilde, 6] . F];
W = Simplify[A . B . Inverse[A] . Inverse[B]];

\end{lstlisting}

\vspace{0.5cm}
\noindent\textbf{Acknowledgment. }
S.X.C is partly supported by NSF grant CCF-2006667, Quantum Science Center (led by ORNL), and ARO MURI.

\bibliographystyle{unsrt}
\bibliography{universal_bib}

\begin{thebibliography}{10}

\bibitem{wen1990topological}
Xiao-Gang Wen.
\newblock Topological orders in rigid states.
\newblock {\em International Journal of Modern Physics B}, 4(02):239--271,
  1990.

\bibitem{freedman2002modular}
Michael~H Freedman, Michael Larsen, and Zhenghan Wang.
\newblock A modular functor which is universal for quantum computation.
\newblock {\em Communications in Mathematical Physics}, 227:605--622, 2002.

\bibitem{kitaev2003fault}
A~Yu Kitaev.
\newblock Fault-tolerant quantum computation by anyons.
\newblock {\em Annals of physics}, 303(1):2--30, 2003.

\bibitem{nayak2008non}
Chetan Nayak, Steven~H Simon, Ady Stern, Michael Freedman, and Sankar~Das
  Sarma.
\newblock Non-abelian anyons and topological quantum computation.
\newblock {\em Reviews of Modern Physics}, 80(3):1083, 2008.

\bibitem{wang2010topological}
Zhenghan Wang.
\newblock {\em Topological quantum computation}.
\newblock Number 112. American Mathematical Soc., 2010.

\bibitem{aghaee2024interferometric}
Morteza Aghaee, Alejandro~Alcaraz Ramirez, Zulfi Alam, Rizwan Ali, Mariusz
  Andrzejczuk, Andrey Antipov, Mikhail Astafev, Amin Barzegar, Bela Bauer,
  Jonathan Becker, et~al.
\newblock Interferometric single-shot parity measurement in an {InAs-Al} hybrid
  device.
\newblock {\em arXiv:2401.09549}, 2024.

\bibitem{aghaee2023inas}
Morteza Aghaee, Arun Akkala, Zulfi Alam, Rizwan Ali, Alejandro Alcaraz~Ramirez,
  Mariusz Andrzejczuk, Andrey~E Antipov, Pavel Aseev, Mikhail Astafev, Bela
  Bauer, et~al.
\newblock {InAs-Al} hybrid devices passing the topological gap protocol.
\newblock {\em Physical Review B}, 107(24):245423, 2023.

\bibitem{minev2024realizing}
Zlatko~K Minev, Khadijeh Najafi, Swarnadeep Majumder, Juven Wang, Ady Stern,
  Eun-Ah Kim, Chao-Ming Jian, and Guanyu Zhu.
\newblock Realizing string-net condensation: {F}ibonacci anyon braiding for
  universal gates and sampling chromatic polynomials.
\newblock {\em arXiv:2406.12820}, 2024.

\bibitem{nakamura2020direct}
James Nakamura, Shuang Liang, Geoffrey~C Gardner, and Michael~J Manfra.
\newblock Direct observation of anyonic braiding statistics.
\newblock {\em Nature Physics}, 16(9):931--936, 2020.

\bibitem{xu2024non}
Shibo Xu, Zheng-Zhi Sun, Ke~Wang, Hekang Li, Zitian Zhu, Hang Dong, Jinfeng
  Deng, Xu~Zhang, Jiachen Chen, Yaozu Wu, et~al.
\newblock Non-abelian braiding of {F}ibonacci anyons with a superconducting
  processor.
\newblock {\em Nature Physics}, 20(9):1469--1475, 2024.

\bibitem{wu2023simulating}
Yong-Shi Wu.
\newblock Simulating the hadamard gate in the {F}ibonacci disk code for
  universal topological quantum computation.
\newblock {\em The Innovation}, 4(6), 2023.

\bibitem{zhan2022universal}
Ye-Min Zhan, Yu-Ge Chen, Bin Chen, Ziqiang Wang, Yue Yu, and Xi~Luo.
\newblock Universal topological quantum computation with strongly correlated
  {M}ajorana edge modes.
\newblock {\em New Journal of Physics}, 24(4):043009, 2022.

\bibitem{witten1989quantum}
Edward Witten.
\newblock Quantum field theory and the jones polynomial.
\newblock {\em Communications in Mathematical Physics}, 121(3):351--399, 1989.

\bibitem{freedman2002two}
Michael~H Freedman, Michael~J Larsen, and Zhenghan Wang.
\newblock The two-eigenvalue problem and density of {J}ones representation of
  braid groups.
\newblock {\em Communications in Mathematical Physics}, 228:177--199, 2002.

\bibitem{cui2019search}
Shawn~X Cui, Kevin~T Tian, Jennifer~F Vasquez, Zhenghan Wang, and Helen~M Wong.
\newblock The search for leakage-free entangling fibonacci braiding gates.
\newblock {\em Journal of Physics A: Mathematical and Theoretical},
  52(45):455301, 2019.

\bibitem{bruguieres2006double}
Alain Brugui{\`e}res.
\newblock Double braidings, twists and tangle invariants.
\newblock {\em Journal of Pure and Applied Algebra}, 204(1):170--194, 2006.

\bibitem{liptrap2010hypergroups}
Jesse Liptrap.
\newblock {\em From Hypergroups to Anyonic Twines}.
\newblock PhD thesis, UNIVERSITY OF CALIFORNIA Santa Barbara, 2010.

\bibitem{bonderson2007non}
Parsa~Hassan Bonderson.
\newblock {\em Non-Abelian Anyons and Interferometry}.
\newblock PhD thesis, California Institute of Technology, 2007.

\bibitem{simon2006topological}
Steven~H Simon, Nicholas~E Bonesteel, Michael~H Freedman, N~Petrovic, and Layla
  Hormozi.
\newblock Topological quantum computing with only one mobile quasiparticle.
\newblock {\em Physical review letters}, 96(7):070503, 2006.

\bibitem{cui2015universal}
Shawn~X Cui and Zhenghan Wang.
\newblock Universal quantum computation with metaplectic anyons.
\newblock {\em Journal of Mathematical Physics}, 56(3), 2015.

\bibitem{rowell2018mathematics}
Eric Rowell and Zhenghan Wang.
\newblock Mathematics of topological quantum computing.
\newblock {\em Bulletin of the American Mathematical Society}, 55(2):183--238,
  2018.

\bibitem{wess1971consequences}
Julius Wess and Bruno Zumino.
\newblock Consequences of anomalous ward identities.
\newblock {\em Physics Letters B}, 37(1):95--97, 1971.

\bibitem{witten1983global}
Edward Witten.
\newblock Global aspects of current algebra.
\newblock {\em Nuclear Physics B}, 223(2):422--432, 1983.

\bibitem{turaev2010quantum}
Vladimir~G Turaev.
\newblock {\em Quantum invariants of knots and 3-manifolds}.
\newblock de Gruyter, 2010.

\bibitem{kitaev1997quantum}
A~Yu Kitaev.
\newblock Quantum computations: algorithms and error correction.
\newblock {\em Russian Mathematical Surveys}, 52(6):1191, 1997.

\bibitem{aharonov2008fault}
Dorit Aharonov and Michael Ben-Or.
\newblock Fault-tolerant quantum computation with constant error rate.
\newblock {\em SIAM Journal on Computing}, 38(4):1207, 2008.

\bibitem{conway1976trigonometric}
John Conway and A~Jones.
\newblock Trigonometric diophantine equations (on vanishing sums of roots of
  unity).
\newblock {\em Acta arithmetica}, 30:229--240, 1976.

\bibitem{cui2018lecture}
Shawn~X Cui.
\newblock Topological quantum computation.
\newblock \url{https://www.math.purdue.edu/~cui177/Lecture_Combined.pdf}, 2018.

\end{thebibliography}

\end{document}